\documentclass[aip,rsi,reprint,graphicx,preprintnumbers]{revtex4-1}

\usepackage{graphicx}

\usepackage{siunitx}
\usepackage[version=4]{mhchem}
\usepackage{dcolumn}
\usepackage{subcaption}
\usepackage[precision=2, unit=mm]{lengthconvert}

\usepackage{hyperref}

\begin{document}

\preprint{\shortstack[r]{This article appeared in Rev. Sci. Instrum. 97, 053307 (2026) and may be found at \url{https://doi.org/10.1063/5.0313157}. \\ Copyright 2026 Authors. This author-accepted manuscript is distributed under a CC BY 4.0 International License.}}

\title{Time-domain anode-decoupling co-design for a floating  microchannel plate detector readout}



\author{Robin F.\ Bonny}
\email{robin.bonny@unibe.ch}

\author{Lorenzo Obersnel}
\author{Martin Rubin}
\author{André Galli}
\author{Peter Wurz}
\author{Rico G.\ Fausch}

\affiliation{Space Research and Planetary Sciences, Physics Institute, University of Bern, Switzerland}


\date{28 May 2026}

\begin{abstract}

We present a microchannel plate (MCP) detector for compact time-of-flight mass spectrometers (TOF-MS) that jointly optimizes the anode geometry and high-voltage AC-decoupling network for electrically floating operation. Undershoot-driven baseline artifacts and pulse broadening are addressed by a time-domain co-design of the anode geometry and decoupling network. The design is validated through a staged workflow that combines full-wave electromagnetic simulations, vector network analyzer measurements, circuit-level transient models, and end-to-end mass spectra. The resulting planar circular patch anode with anode-proximal decoupling confines fields, preserves peak amplitude, and suppresses post-pulse energy, leading to fast settling and minimal baseline wander. We show that the effective high-pass corner set by the decoupling capacitance directly governs undershoot decay and baseline recovery. Measurements in a representative TOF-MS test setup demonstrate waveguide-level pulse fidelity at a fraction of the mass and volume of heritage waveguide-based detectors, with residual ripples in the measured response originating from downstream cable and digitizer terminations rather than the detector itself. By limiting detector-induced temporal broadening and inter-peak baseline coupling, the design supports high mass resolution and dynamic range in miniaturized TOF-MS architectures. Variants of this planar flight-ready architecture are being implemented in several next-generation spaceborne TOF-MS instruments currently under development at the University of Bern.

\end{abstract}

\maketitle

\section{Introduction}

\subsection{Time-of-flight mass spectrometry}

Time-of-flight mass spectrometers (TOF-MS) have been instruments of choice for planetary exploration for multiple decades due to their fast acquisition rates, broad mass ranges, extensive dynamic range, and spectral continuity. \cite{arevalo_jr_mass_2020, yang_review_2025, vazquez_situ_2021, ren_review_2018} A TOF-MS operates on the principle that ions are extracted and accelerated through a well-defined potential, such that, to first order, they acquire a common kinetic energy and separate according to their mass-to-charge ratio ($m/z$) during their flight. Mass spectra are acquired through pulsed operations: Ions with different $m/z$ will arrive at a detector at distinct times \cite{wiley_timeflight_1955} according to the relation \(t\propto\sqrt{m/z}\) and with respect to their extraction time. Therefore, precise measurements of the ion arrival times on the detector are critical to obtain high mass resolution.

\subsection{Detection limitations}

Since spaceborne mass spectrometers are allocated limited size, weight, and power budgets, there is an extrinsic need to miniaturize instruments. When downsizing time-of-flight mass spectrometers, the time of flight decreases and the mass resolution is reduced, as in \(R=m/\Delta m = t/(2\Delta t)\). The detector has a considerable influence on the mass resolution of a TOF instrument. The time-resolving power ($\Delta t$) of a TOF-MS can be approximated as the quadrature sum of independent timing contributions; including ion-source pulse width, ion-optical aberrations, detector response, and electronics jitter. Among these, the detector contribution is set by its single-event pulse width, defined here as the full width at half-maximum (FWHM). To comply with the required ion-optical resolving powers of spaceborne instruments, for example, the Neutral Gas and Ion Mass Spectrometer (NIM) built for the Jupiter Icy Moons Explorer (JUICE) mission, \cite{fohn_description_2021} the detector pulse width ideally equals about \qty{500}{\ps} or less.

When coadding multiple peaks into a histogram, the noise floor decreases and low-amplitude systematic artifacts after the main peak become apparent. Some features can be attributed to impedance mismatch, while others are characteristic of high-voltage decoupling capacitors in the signal transmission line that serve as AC-decoupling elements toward front-end electronics. The AC-decoupling stage forms an effective high-pass response: Low-frequency components of the MCP-induced pulse are attenuated or blocked, so the coupled waveform must recover toward a zero-mean baseline over time. This produces a characteristic negative tail (i.e.\ undershoot) following the main pulse, whose decay is governed by the effective high-pass corner frequency.

In practice, this baseline-recovery waveform is further shaped by unavoidable effects: Parasitic capacitances (anode-to-ground, pads and vias, downstream input capacitances) add shunt loading that reduces effective bandwidth and mainly impacts the falling edge and pulse tail, while parasitic inductances (vias, interconnects and capacitor nonidealities) can resonate with these capacitances and manifest as ringing. The combination of these artifacts compromises the detection of small peaks following large ones, effectively reducing the dynamic range of the TOF-MS when closely spaced peaks are present. For example, analyzing \ce{Xe} isotopes \cite{fausch_direct_2022} or \ce{Rb}/\ce{Sr} isotopes \cite{levine_dating_2023} becomes challenging, as such isotope patterns often result in small minor isotopologue peaks appearing in close succession after the large peaks of the main isotopologues. Moreover, this undershoot can complicate fully automated analysis of mass spectra, \cite{meyer_fully_2017} since the baseline of a given peak would depend on the preceding ones and vice versa.

These detector-specific limitations are central to the requirements of the compact TOF-MS architectures, including the Chemistry, Organics, and Dating Experiment (CODEX) instrument, currently under development and which serves as the primary testbed for the design and validation of the work presented in this paper. CODEX is part of the Dating an Irregular Mare Patch with a Lunar Explorer (DIMPLE) payload aboard a Commercial Lunar Payload Services (CLPS) lander. \cite{Fausch2026, Wurz2026}

\subsection{Detector architecture}

Spaceborne TOF-MS detectors often employ microchannel plates (MCPs), which convert an incident ion into an avalanche of electrons, collected on a downstream anode. \cite{wiza_microchannel_1979} A single ion interaction produces a current impulse whose temporal shape is approximately Gaussian. \cite{wiza_microchannel_1979} Furthermore, TOF-MS detector systems routinely operate across an extremely wide dynamic range, spanning single-ion detection to ion bunches containing millions of ions, depending on source mode and operating conditions. \cite{hohl_mass_1999} When many ions arrive within a narrow time window, the resulting ion-packet peak reflects both the intrinsic MCP response and the time dispersion introduced by the ion-optical system. Their relative contributions depend on the specific energy and mass range of the measurement. In addition, the integrated peak area of the detected signal is proportional to the number of ions in the packet. \cite{wurz_fast_1996}

The MCP input side (MCP-front) is held at or slightly more negative than the ion-optical reference potential of the mass analyzer. The MCP output side (MCP-back) is biased up to a few kilovolts more positive than this level to set the MCP gain, which may need to be adjusted over the lifetime of the MCP. To efficiently collect the electrons exiting the MCP, the anode surface is biased an additional \qtyrange{100}{200}{\V} more positive than MCP-back, providing sufficient post-acceleration for fast electron collection and a short output pulse. Because the readout electronics are usually referenced to true ground, whereas the MCP and anode operate at several kilovolts of negative potential, the anode must be electrically floating relative to the readout chain. This floating configuration also decouples the detector from the ion-optical reference potential and therefore allows the drift voltage to be selected freely. The signal is AC coupled into the transmission line, and to obtain optimal signal fidelity, the AC-coupling capacitors must be located as close as possible to the anode surface. \cite{wurz_fast_1996} 

The trade-off whether to implement an electrically floating anode was studied for the NIM instrument. \cite{lasi_decisions_2020} Floating the anode enables more degrees of freedom in the design and operation of the ion-optical system and MCP bias at the cost of requiring an anode-proximal AC-decoupling stage. The NIM trade-off study concluded that these advantages outweighed the added complexity. \cite{lasi_decisions_2020} Grounded-anode alternatives include segmented circular microstrip readouts \cite{riedo_high-speed_2017} or fully impedance-matched transmission-line anodes directly connected to front-end electronics. \cite{fausch_advancing2_2025}

\begin{figure*}
    \centering
    \includegraphics[width=.9\textwidth]{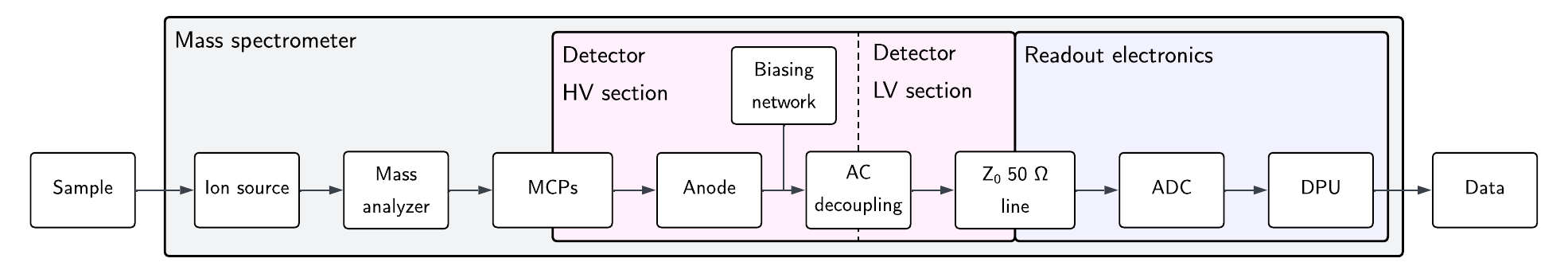}
    \caption{\label{fig:detec_block} Conceptual overview of the floating MCP detector readout architecture and its separation into floating high-voltage and ground-referenced low-voltage domains. The anode signal is transferred across the isolation boundary by coupling capacitors into a \qty{50}{\ohm} transmission-line and readout environment. ADC denotes the analog-to-digital converter and DPU the data processing unit.}
\end{figure*}

\begin{figure}
    \centering
    \includegraphics[width=\columnwidth]{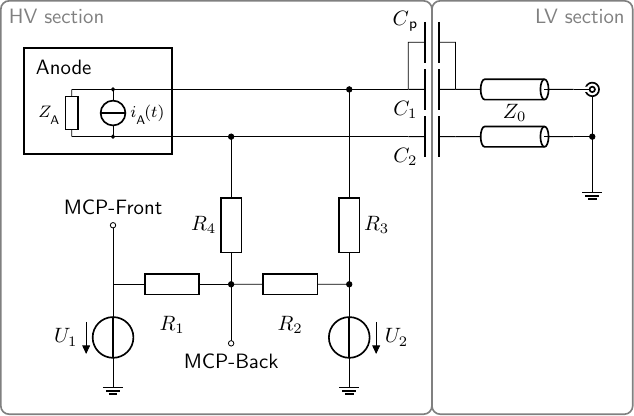}
    \caption{\label{fig:final_circ} Lumped-element equivalent circuit of the floating MCP anode and AC-decoupling network, separated into a high-voltage (HV, floating) and a low-voltage (LV, ground-referenced)  section. $U_1$ and $U_2$ denote the externally applied MCP-Front and Anode potentials, respectively; the MCP-Back potential is established by the internal bias network between these nodes.}
\end{figure}

All detector configurations evaluated in this work assume a floating anode with a high-voltage AC-decoupling capacitor for bias voltages up to \qty{-2}{\kV}. A block-level overview of this architecture is shown in Fig.~\ref{fig:detec_block} and the detailed electrical layout is presented in Fig.~\ref{fig:final_circ}.

\subsection{Reference designs and data}

\begin{figure}
    \centering
    \includegraphics[width=.7\columnwidth]{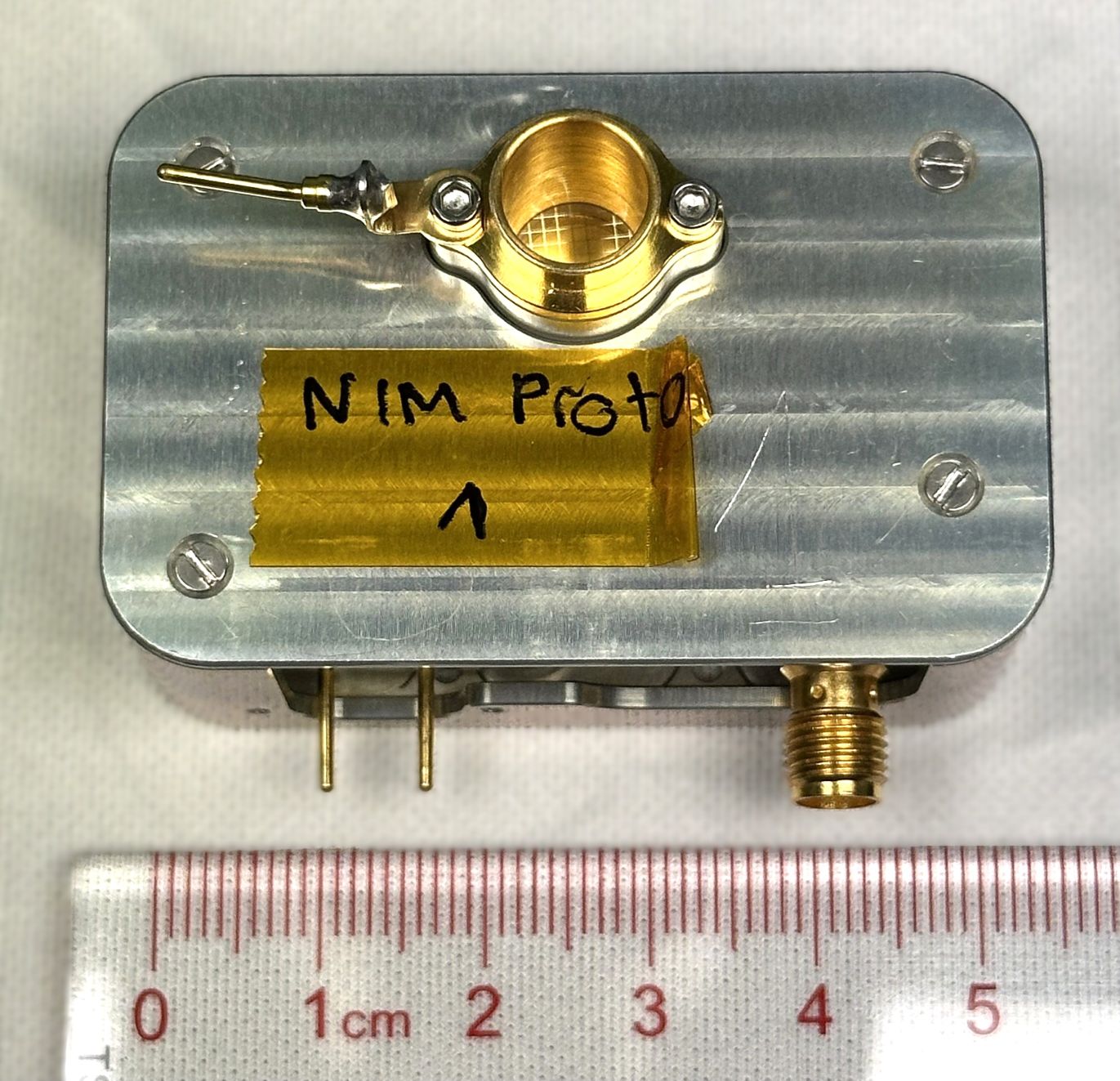}
    \caption{\label{fig:proto} Laboratory test detector assembly used for the end-to-end measurements reported in this work. This external mechanical package (housing, ion aperture, connectors) is identical for the NIM and CODEX configurations.}
\end{figure}

To contextualize the performance of our developed detector, hereafter denoted as the \emph{CODEX} detector, \cite{Fausch2026} we studied mass spectra of comparable detector systems. These include two prototype detectors of NIM with different anode geometries, \cite{lasi_decisions_2020} very closely comparable in terms of size and design requirements to our proposed solution. These NIM detectors are hereafter referred to as \emph{NIM-A} and \emph{NIM-B}. Specifically, \emph{NIM-A} employs an anode of circular shape, whereas \emph{NIM-B} features a spiral-type anode. A view of the laboratory test detector assembly is shown in Fig.~\ref{fig:proto}; the NIM and CODEX configurations share the same external housing and interfaces, differing only in the internal anode and the implementation of the AC-coupling.

As a reference, we also compared our measurements with a new mass spectrometer currently being developed at the University of Bern for an atmospheric Uranus entry probe, \cite{vorburger_mass_2024} which inherits the waveguide-based detector concept of \citeauthor{wurz_fast_1996}, \cite{wurz_fast_1996} and builds on the instrument architecture of \citeauthor{abplanalp_neutral_2009} \cite{abplanalp_neutral_2009} Because this detector design achieves an almost ideal impedance profile, its pulse shape closely approaches the ground-truth impulse response and thus serves as a high-fidelity comparison standard. We refer to this detector as \emph{UOP}, after the Uranus Orbiter and Probe mission concept for which this instrument is being developed. A related waveguide-based MCP detector design, building on the concepts of \citeauthor{wurz_fast_1996}, is also used in the Mass Spectrometer for Planetary Exploration (MASPEX) on the Europa Clipper spacecraft. \cite{waite_maspex-europa_2024}

\section{Methods}

\subsection{Signal-path abstraction}

In the context of the CODEX instrument, the detector output is digitized at a sampling rate of 2.5 gigasamples per second (\unit{\giga Sa \per\s}) after analog filtering to prevent aliasing effects. This sampling rate is sufficiently high to digitize the detector output with acceptable distortion at the expected temporal pulse widths of the electron avalanche peaks, as validated in similar contexts. \cite{lasi_decisions_2020, fohn_description_2021, meyer_fully_2017, riedo_high-speed_2017, abplanalp_neutral_2009}

For Gaussian peak shapes with a given $t\textsubscript{FWHM}$, the frequency spectrum is also Gaussian. As a general rule, the bandwidth scales approximately with $1/t\textsubscript{FWHM}$. For $t\textsubscript{FWHM}$ in the few-hundred-picosecond range, the corresponding frequency content extends well into the gigahertz bandwidth regime, with most of the spectral energy contained below \qty{2}{\GHz}, which we therefore adopt as the upper design frequency for the analog detector response. In practice, however, the usable bandwidth of the digitized signal is limited by the \qty{2.5}{\giga Sa\per\s} sampling rate and the anti-aliasing filter, which constrain the effective bandwidth to below the Nyquist frequency of \qty{1.25}{\giga\hertz}. Therefore, using the Gaussian time--frequency scaling introduced above as a rule of thumb, this setup is expected to resolve Gaussian peaks only down to roughly \qty{800}{\ps}. Expressed instead in terms of rise time, the same bandwidth corresponds to a lower limit on the order of a few \qty{100}{\ps}, using \(t_r \approx 1/(3\cdot\mathrm{BW})\). These estimates are approximate and reflect different pulse-width conventions, but are broadly consistent with experimentally observed \qty{500}{\ps}-class pulses.

At these frequencies, lumped-element circuit theory becomes insufficient and must be complemented by transmission line theory, as the geometric dimensions on the detector reach considerable fractions of the effective wavelength in the substrate. Equation~(\ref{eq:lambda}) computes this value for an assumed velocity factor of \num{0.8}. Typically, we assume that distributed effects of impedance mismatch start to become noticeable when the relevant dimensions approach or exceed $\lambda/10$.
\begin{equation}
    \frac{\lambda}{10} = \frac{1}{10}\cdot\frac{c}{f} \approx \frac{0.8\,c_0}{10\cdot\qty{2}{\GHz}} \approx \qty{1.2}{\cm}
    \label{eq:lambda}
\end{equation}

Therefore, we model the signal paths on the detector, specifically the coplanar waveguides (CPWGs), as transmission lines with characteristic impedances $Z_0$. Electrically floating MCP detectors have evolved toward geometries that approximate a constant \qty{50}{\ohm} line impedance throughout the analog signal path up to the readout, resulting in microstrip-inspired anode structures. \cite{fohn_description_2021, riedo_high-speed_2017, schletti_fast_2001}

\subsubsection{Anode: Low-pass limit}

The detector anode can be approximated using models developed for circular microstrip patch structures, which provide analytic estimates of the corresponding resonant frequencies. The zeroth-order resonant frequency of a circular patch on a thin substrate is given by Eq.~(\ref{eq:patchf}) derived by \citeauthor{shen_resonant_1977}, \cite{shen_resonant_1977} with $c_0$ the speed of light in free space, $a$ the radius of the patch, and $\varepsilon_r$ the relative permittivity of the substrate, under the assumption that \(\mu\approx\mu_0\). When the operating frequency range approaches such resonances, the anode exhibits ringing and strong dispersive distortion. Therefore, the lowest resonance defines an upper bound for the usable frequency range of circular anode designs.
\begin{equation}
    f^{(0)} = \frac{1.841}{2\pi a\sqrt{\mu\varepsilon}} = \frac{1.841 c_0}{2\pi a \sqrt{\varepsilon_r}}
    \label{eq:patchf}
\end{equation}

For a circular patch on a standard FR-4 substrate (fiberglass-epoxy material) with a typical permittivity of \(\varepsilon_r\approx4.4\) and a radius of $a=\qty{5}{\mm}$, this leads to a maximum frequency of \qty{8.4}{\GHz}.

\subsubsection{Decoupling: High-pass limit}

The lower frequency bound of the detector system is set by the time constant $\tau\textsubscript{eff}$ of the overall effective resistance $R\textsubscript{eff}$ and capacitance $C\textsubscript{eff}$ of the circuit, as shown in Eq.~(\ref{eq:rc}). Depending on which parts of the detector system are connected to the ground potential and how current return paths are routed external to the detector printed circuit board (PCB), the value of $\tau\textsubscript{eff}$ depends on the configuration and differs between setups. A representative value of $\tau\textsubscript{eff}$ is stated for the developed CODEX implementation under the final design.
\begin{equation}
    f_c = \frac{1}{2\pi\tau\textsubscript{eff}} = \frac{1}{2\pi R\textsubscript{eff}C\textsubscript{eff}}
    \label{eq:rc}
\end{equation}

To avoid introducing pronounced undershoot and baseline distortion, we choose $f_c$ sufficiently below the pulse bandwidth (i.e., $f_c \ll 1/t\textsubscript{FWHM}$), so that AC coupling mainly sets the post-pulse recovery rather than distorting the main peak.

\subsection{Development workflow}

To address the complementarity and interaction between the anode and the decoupling network, we followed a staged development workflow that converged toward the final optimized detector design:

\begin{enumerate}
    \item Full-wave simulation of candidate anode geometries independently of the downstream circuitry;
    \item Design, fabrication, and Vector Network Analysis (VNA) characterization of a prototype PCB implementing the validated geometries;
    \item Circuit-level modeling to reproduce the measured $S$-parameters and to simulate the corresponding time-domain response; and
    \item End-to-end mass spectrometric measurements using the assembled detector PCB within a TOF-MS test setup.
\end{enumerate}

Some steps are particularly devised to study either individual components or the integrated design. For example, full-wave simulations yield the best insights into the anode behavior isolated from the contributions of the rest of the circuit, whereas the VNA measurements are crucial for the synthesis of the decoupling network. The mass spectrometric measurements reveal the interplay between all elements.

The individual steps and their corresponding methods are laid out in more detail in the following sections.

\subsubsection{Full-wave simulation}

Given the gigahertz-scale bandwidth and transmission-line behavior of the detector anode and signal paths, a full-wave electromagnetic analysis is required to capture the effects of geometry and impedance profiles. We therefore analyzed candidate anode geometries using \emph{Ansys HFSS} (High-Frequency Structure Simulator) and subsequently verified the simulation results with laboratory prototyping measurements.

\begin{figure}
\centering
\begin{subfigure}{.5\textwidth}
  \centering
  \includegraphics[width=.95\columnwidth]{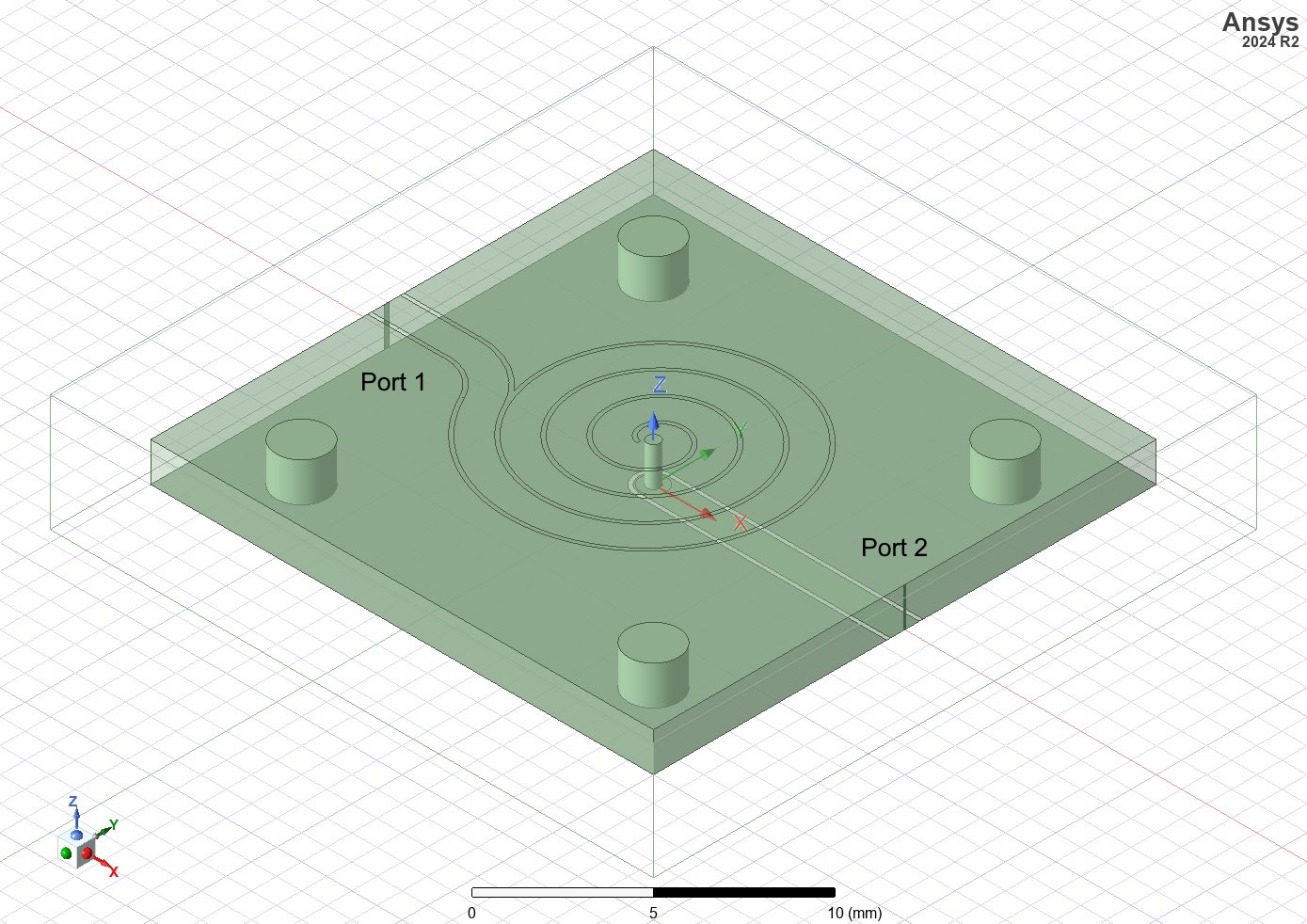}
  \caption{Spiral anode geometry (NIM)}
\end{subfigure}
\begin{subfigure}{.5\textwidth}
  \centering
  \includegraphics[width=.95\columnwidth]{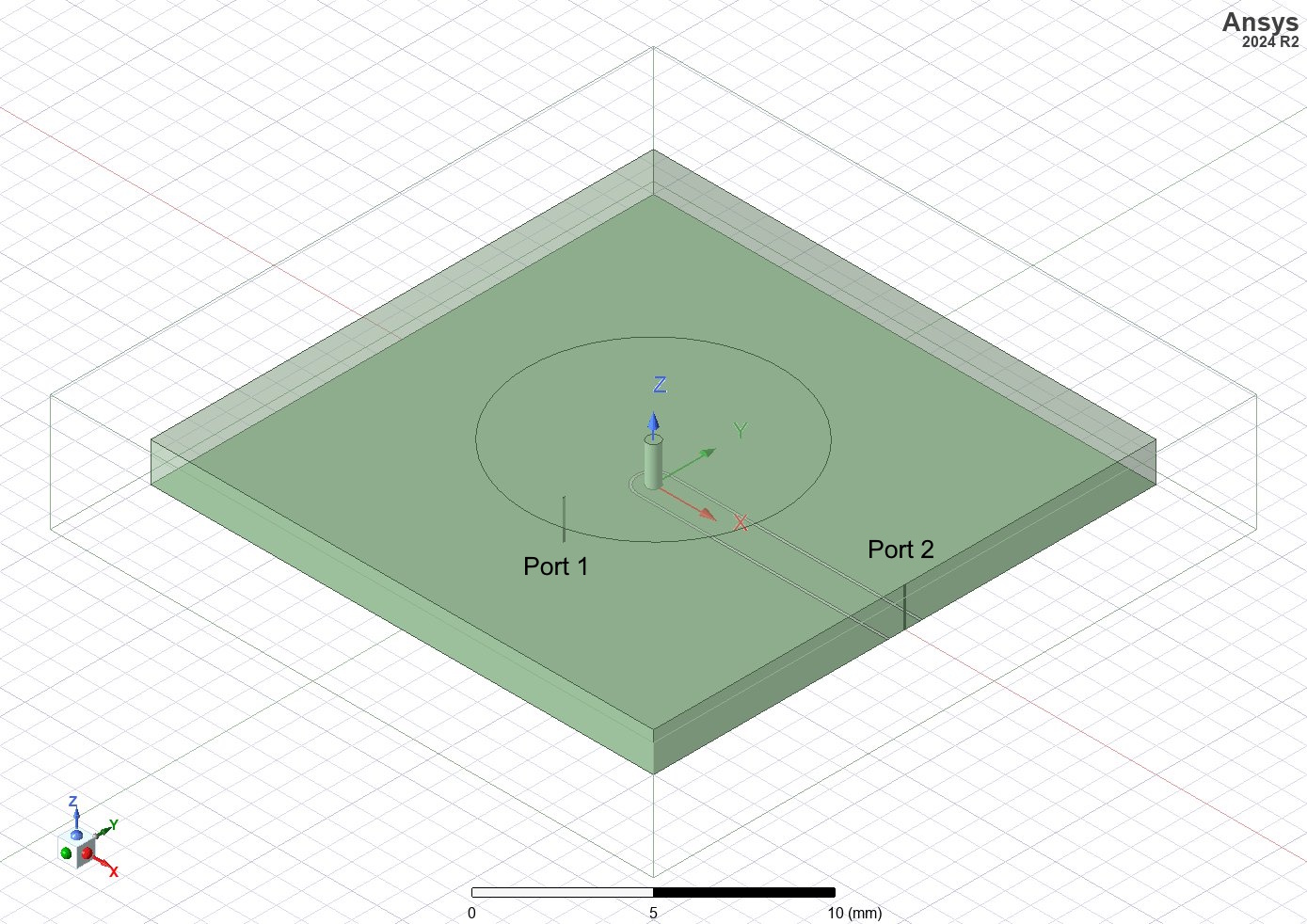}
  \caption{Circular patch anode geometry (CODEX)}
\end{subfigure}
\caption{\label{fig:hfss} HFSS full-wave models used for electromagnetic characterization of the anode signal path, configured as two-port problems. Scale bar and port annotations shown.}
\end{figure}

We simulated the frequency response of several anode geometries, specifically the spiral layout used in NIM and a circular patch variant. Simulations were configured as two-port problems: One port being the interface of the anode with the CPWG leading to the decoupling capacitor, the other being a ``virtual'' port located on the anode structure, emulating a localized electron impact along the signal path. A visualization of the HFSS full-wave models for spiral and circular patch anode geometries with microstrip readout lines is shown in Fig.~\ref{fig:hfss}.

Frequency-domain simulations were set up as HFSS solution types in modal network analysis with a solving frequency of \qty{5}{\GHz}. Time-domain simulations employed the transient solution type with smooth (i.e., Gaussian) pulse excitation of \qty{1}{\ns} full envelope duration, corresponding to an FWHM of \qty{0.32}{\ns}.

\subsubsection{Electrical performance}

With insights on anode shapes from the full-wave simulations, we designed prototype detector PCBs that also include the AC-decoupling stage. Electrical benchtop measurements can be performed to better understand time and frequency behavior before advancing to more sophisticated testing techniques. These PCBs were implemented in FR-4 class laminate, to meet the short turnaround times within the staged development workflow.

We describe the readout as a linear two-port using scattering parameters ($S$-parameters), where \(S_{ij} = b_i/a_j\) is the ratio of reflected to incident complex wave parameters at ports $i$ and $j$, referenced to a given port impedance (typically \qty{50}{\ohm}). \cite{kurokawa_power_1965} Of most interest here are the transmission terms $S_{21}$ and $S_{12}$, which quantify forward and reverse signal transfer through the network.

These parameters can be obtained by introducing a second ``virtual'' port, similar to what has been done for full-wave simulation. For this purpose, we soldered a coaxial pigtail to the CPWG on the anode backside to characterize transmission from this point on the transmission line through the decoupling and via the regular readout. In this two-port configuration, the anode is not part of the through path but appears as a shunt stub impedance at the anode-side node and, therefore, still loads the measured response. Using such setups for different hardware iterations, we measured $S$-parameters using a VNA for multiple combinations of component values in the biasing and decoupling network. Specifically, VNA measurements were acquired in the frequency range of \qty{5}{\kHz} to \qty{3}{\GHz} after full two-port electronic UOSM (unknown-open-short-match) calibration. 

For the VNA measurements, $\tau\textsubscript{eff}$ can only be determined based on the contribution of $C_1$, as the VNA ties both sides of the reference plane to ground, and therefore no current passes through $C_2$. Nevertheless, this allows for the characterization of parasitic capacitance $C\textsubscript{p}$ in parallel with $C_1$.

\subsubsection{Circuit-level transient model}

The frequency-domain $S$-parameter data obtained from the VNA measurements can be converted into time-domain behavior through rational fitting. Since the final data product of the TOF-MS is time-based, this step provides important insights into the expected pulse shapes of the final setup. To better understand the contribution of individual circuit elements, we implemented two complementary circuit-level transient modeling approaches:
\begin{itemize}
    \item A physics-based equivalent lumped-element circuit model based on the transmission line parameters and component values of the prototype PCB; and
    \item A black-box model which embeds the measured two-port $S$-parameters directly into a time-domain simulation.
\end{itemize}

\begin{figure}
    \centering
    \includegraphics[width=\columnwidth]{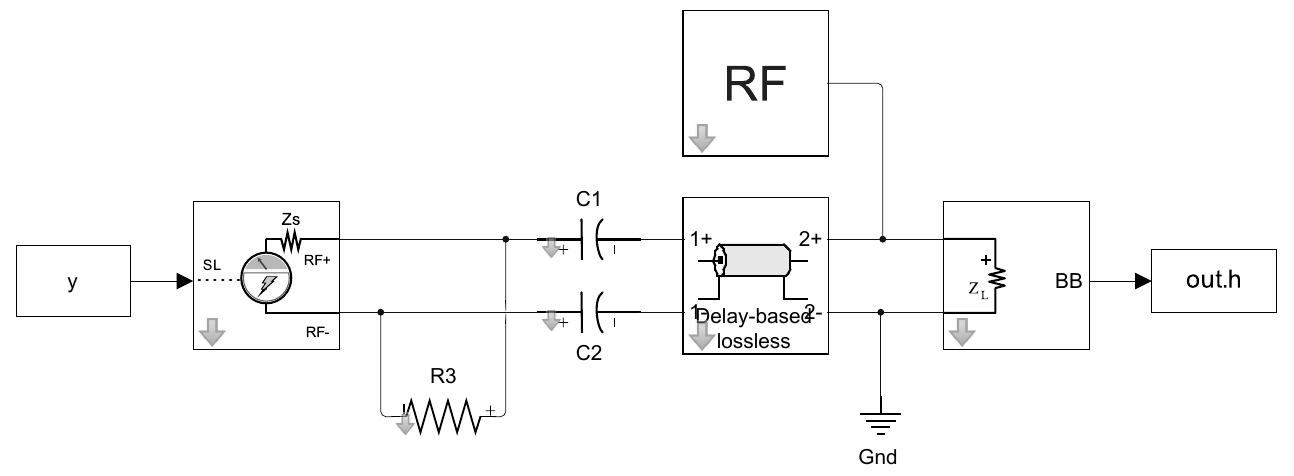}
    \caption{\label{fig:simlink_syn} Simulink circuit-envelope time-domain model with ideal lumped elements used to evaluate baseline restoration and undershoot in the floating-anode decoupling network. The model is driven and terminated with \(Z\textsubscript{S} = Z\textsubscript{L} = \qty{50}{\ohm}\); the baseline-return element is \(R_3 = \qty{1}{\Mohm}\).}
\end{figure}

To study the transient response, only the small-signal model of the full circuit previously shown in Fig.~\ref{fig:final_circ} is considered. We implemented this circuit in Simulink using the RF blockset as a circuit-envelope simulation for broadband analysis as shown in Fig.~\ref{fig:simlink_syn}. The excitation pulse used in this transient model differs from the intrinsic MCP single-event pulse width: Because the goal is to reproduce the electrical transient behavior of the detector PCB under representative test conditions, we employed Gaussian current impulses with an FWHM of \qty{2.5}{\ns}, matching the pulse widths observed in subsequent end-to-end measurements in the \emph{Messkammer für Flugzeit-Instrumente und Time-Of-Flight} (MEFISTO), a custom TOF-MS test setup \cite{abplanalp_neutral_2009, fausch_advancing2_2025} housed in a vacuum calibration chamber of the same name. \cite{wurz_new_1998, marti_calibration_2001} These simulations assume a \qty{50}{\ohm} source and load (\( Z\textsubscript{S} = Z\textsubscript{L} = \qty{50}{\ohm}\)) as well as a baseline-return resistor \(R_3 = \qty{1}{\Mohm}\). In this configuration, the observed peak width reflects the convolution of the test ion-optical arrival-time distribution with the detector and readout impulse response; it is not the intrinsic single-event MCP avalanche width. This choice provides a realistic broadband stimulus to evaluate the combined influence of the decoupling network, parasitic elements, and the transmission-line structure.

A key advantage of this circuit-level approach is that neither port of the model is inherently tied to ground, allowing the effective time constant $\tau_\text{eff}$ to include contributions from both capacitors $C_1$ and $C_2$.

\begin{figure}
    \centering
    \includegraphics[width=\columnwidth]{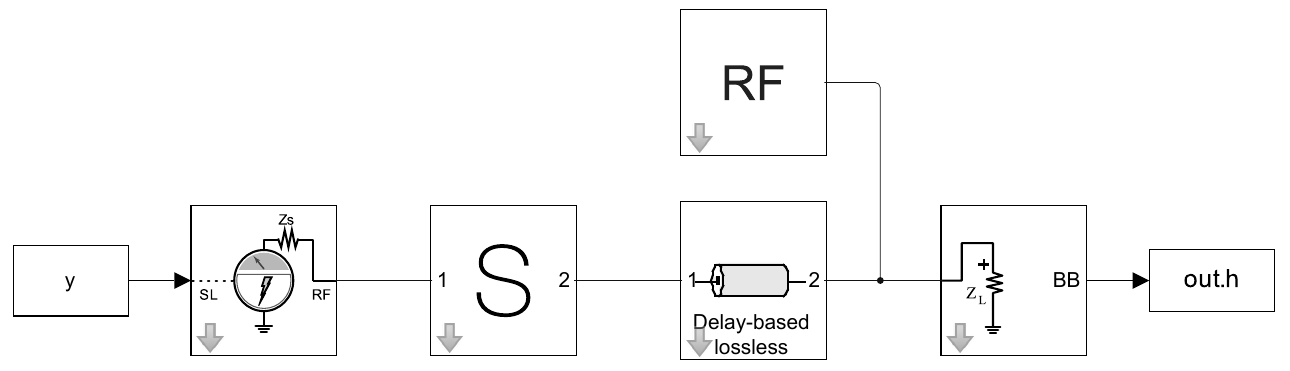}
    \caption{\label{fig:simlink_ref} Simulink circuit-envelope time-domain model in which the anode-interconnect path is represented by measured two-port $S$-parameters from the VNA. The model uses the same \qty{50}{\ohm} source and load environment.}
\end{figure}

The black-box approach extends this method by replacing the analytical component models with the measured $S$-parameters, as shown in Fig.~\ref{fig:simlink_ref}. This enables a validation of the small-signal circuit model against the physical measurements and reveals the combined influence of both capacitors on the transient response. However, both methods exclude contributions from power supplies and external parasitics and therefore provide an accurate electrical response of the detector PCB but not of the fully integrated system.

Certain aspects of the detector behavior, particularly slow baseline-recovery processes governed by large effective time constants, cannot be inferred from frequency-domain data alone. Therefore, time-domain measurements remain essential to validate the full transient response of the detector.

\subsubsection{End-to-end mass spectra}

Once the electrical performance had been validated, we tested the prototype detector in the MEFISTO test setup. This setup allowed us to acquire full end-to-end time-of-flight mass spectra, most importantly including the response of the physical MCP which will inherently be part of the final data acquisition chain. This setup comes with its own limitations: long turnaround times when modifying detector iterations due to pump-down and outgassing times, as well as nonidealities in the readout chain.

End-to-end measurements were performed with the ion-optical assembly of the MEFISTO setup, which differs from that of the final CODEX flight instrument. The intent here is to characterize the detector and readout transient response rather than the fully integrated CODEX-MS system. Accordingly, only representative individual peaks are shown in this work. Full CODEX mass spectra and further instrument-specific details are presented by \citeauthor{Fausch2026} \cite{Fausch2026}

\subsection{Metrics and processing}

\begin{figure*}
    \centering
    \includegraphics[width=.7\textwidth]{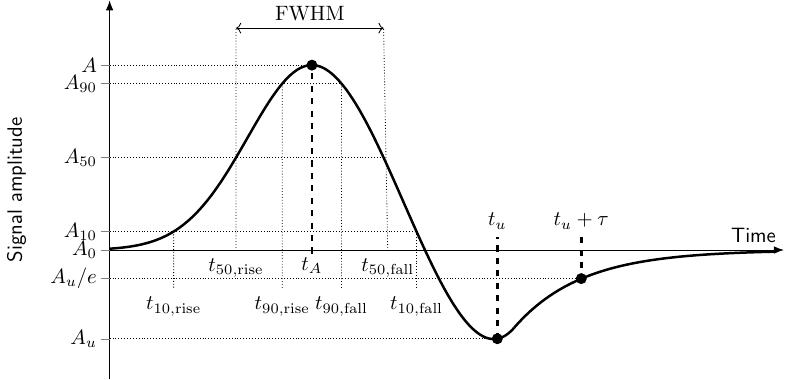}
    \caption{Time-domain pulse metrics derived from IEEE 181 used for quantitative detector responses, illustrated on a representative waveform.}
    \label{fig:metrics}
\end{figure*}

To quantify and compare the performance of the different detector iterations, we used a range of numerical metrics derived and extended from the \emph{IEEE Standard for Transitions, Pulses,
and Related Waveforms} (IEEE Std 181). \cite{noauthor_ieee_2011} A graphic overview of these metrics is shown in Fig.~\ref{fig:metrics} on a simplified signal shape.

\subsubsection{Basic metrics}

Basic metrics are evaluated from the crossing times of the \num{10}, \num{50}, and \qty{90}{\percent} amplitude levels ($t_{10}$, $t_{50}$, $t_{90}$ corresponding to $A_{10}$, $A_{50}$, $A_{90}$) for both the positive- and negative-going transitions of the pulse between the baseline $A_0$ and the peak amplitude $A$ (crossing time $t_A$). The undershoot peak $A_u$ (crossing time $t_u$) is defined as the largest negative excursion following the main positive-going pulse.

\newcommand{\lbl}[1]{\parbox[t]{0.34\columnwidth}{\raggedright\small #1}}

\begin{align}
\lbl{Transition duration\\(positive-going)} &:= t_{10\text{--}90}=t_{90,\mathrm{rise}}-t_{10,\mathrm{rise}} \\
\lbl{Transition duration\\(negative-going)} &:= t_{90\text{--}10}=t_{10,\mathrm{fall}}-t_{90,\mathrm{fall}} \\
\lbl{Pulse duration (FWHM)} &:= t_{50,\mathrm{fall}}-t_{50,\mathrm{rise}} \\
\lbl{Asymmetry index} &:= \frac{t_{10\text{--}90}-t_{90\text{--}10}}{t_{10\text{--}90}+t_{90\text{--}10}} \\
\lbl{Undershoot} &:= -\frac{A\textsubscript{u}-A_0}{A-A_0}
\end{align}

\subsubsection{Advanced metrics}

The transition settling duration is defined as the difference between the settling time $t\textsubscript{settle}$ and the \qty{50}{\percent} crossing time on the falling edge. The settling time itself is defined as the earliest instant at which all subsequent samples fall within a specified tolerance band $E$ around the baseline, as shown in Eq.~(\ref{eq:settle}). This value is adopted here as a post-processing metric to quantify baseline recovery, analogous to standard settling-time definitions used for step responses in linear systems. We report both the \qty{1}{\percent} and \qty{0.1}{\percent} settling durations. 
\begin{equation}
    t\textsubscript{settle} := \inf\big\{ t\geq t_{50,\text{fall}} : \lvert A(u)-A_0 \rvert \leq E\;  \forall u\in[t,t\textsubscript{end}]  \big\}
    \label{eq:settle}
\end{equation}
The tail time constant is defined from the observed exponentially decaying shape of the undershoot peak. Because the undershoot is well approximated by a negative-going exponential recovery toward the baseline, the time constant $\tau$ is obtained from the crossing of the amplitude $A_u/e$ from the absolute value of the maximum undershoot excursion. This time constant corresponds to the characteristic cutoff frequency of the corresponding decay and therefore, in a first-order $RC$ model, to the effective resistance–capacitance product governing the baseline recovery.

\section{Results}

\subsection{Anode shape}

\begin{figure}
    \centering
    \includegraphics[width=\columnwidth]{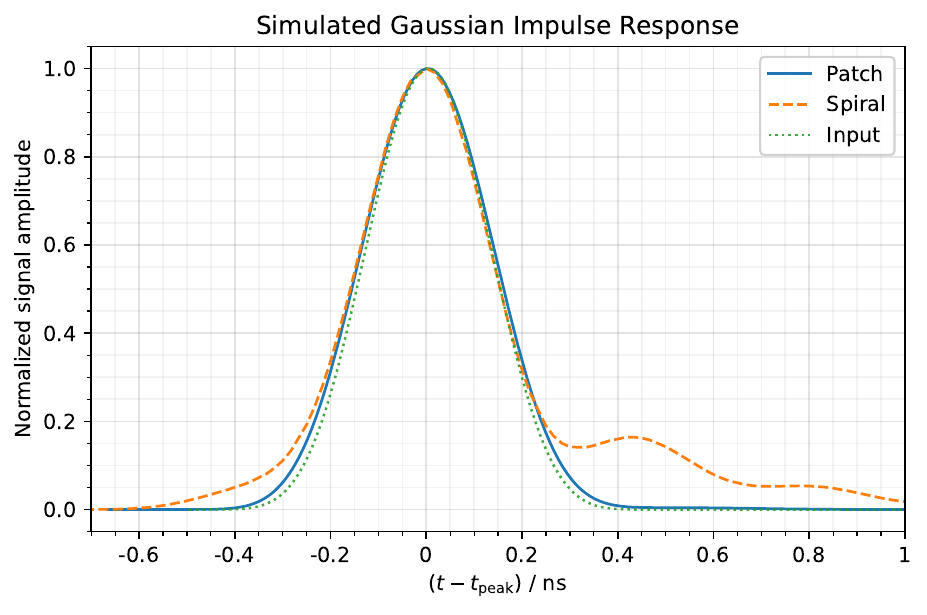}
    \caption{\label{fig:hfss_time} HFSS transient simulation responses of the circular patch (CODEX) and the spiral (NIM) anode geometries to a Gaussian excitation (full envelope \qty{1}{\ns}, FWHM \qty{0.32}{\ns}). Traces are peak-aligned and normalized to compare pulse-shape distortion.}
\end{figure}

The full-wave simulations showed considerable time-domain signal deterioration for the spiral anode shape, contradicting earlier expectations that its CPWG structure would improve signal integrity through better impedance matching. The sub-nanosecond Gaussian excitation used here is a broadband pulse chosen to isolate intrinsic anode dispersion; broader end-to-end peak widths measured in MEFISTO additionally include MCP and acquisition-chain effects and are discussed separately. Fig.~\ref{fig:hfss_time} compares the spiral and circular patch responses after pulse propagation from the ``virtual'' input port to the output. For clarity, the pulses have been aligned in time and normalized in amplitude to isolate shape distortions from propagation delays or losses.

The spiral introduces dispersive effects caused by electromagnetic spillover between the windings and poor field confinement, broadening the overall pulse envelope and introducing secondary peaks. In contrast, the circular patch preserves a nearly Gaussian profile, with only the highest frequencies being cut off due to its low-pass nature, leading to a very slight broadening on the order of tens of picoseconds. From the HFSS modal-network transmission response, the circular patch anode exhibits a \qty{-3}{\dB} bandwidth of approximately $\qty{2.76}{\GHz}$ in $\lvert S_{21} \rvert$, corresponding to a minimum transmitted Gaussian pulse width of \(\mathrm{FWHM}\approx 1/\mathrm{BW}\approx \qty{0.36}{\ns}\) using the time-frequency scaling introduced earlier.

These results indicate that a circular patch anode offers superior signal fidelity at the detector front end compared to the spiral geometry. In addition, the circular layout provides a slightly larger active area for a given radius and avoids the increased propagation delay inherent to long spiral traces. This reduced delay becomes relevant when the anode size is weighed against the achievable timing resolution.

\subsection{Decoupling capacitance and resistance}

VNA measurements allow for an accurate characterization of the frequency contribution of $C_1$, including any parallel parasitics. This characterization was repeated over successive hardware iterations of the newly developed CODEX detector. The corresponding component values and resulting frequency metrics are summarized in Table~\ref{tab:vna_freqs}. While $f_{\text{lo},1}$ and $f_{\text{lo},2}$ are directly determined by the coupling network in the lumped-element sense, $f\textsubscript{hi}$ is reported as an empirical upper-bandwidth metric of the assembled path and is not interpreted as being directly governed by the values of $C_1$ and $C_2$.

For the two-port VNA characterization, port~1 was defined by a coaxial pigtail soldered to the CPWG on the backside of the detector PCB at the anode-side signal node, while port~2 corresponds to the nominal readout connector on the electronics side. Therefore, measurements capture the integrated PCB signal path (launches, vias, traces, and the decoupling and bias network) under \qty{50}{\ohm} source and load conditions. The anode is not in the direct signal through path but appears as a shunt stub impedance at the anode-side node and thus still contributes to the measured response.

The low-pass frequency limit $f\textsubscript{hi}$ is extracted as the \qty{-3}{\dB} bandwidth of the measured $\lvert S_{21} \rvert$ response of the assembled PCB path and therefore captures the combined low-pass behavior of the anode and interconnect geometry as well as any bandwidth limitations introduced by component nonidealities (e.g., ESL and self-resonances of capacitors) and transitions (vias and launches).

The high-pass cutoff frequency $f_{\text{lo},1}$ follows the expression derived in Eq.~(\ref{eq:tau1}). A concise derivation using the $ABCD$ formalism is provided in Appendix~\ref{sec:abcd}. This relationship allows for the quantification of the parasitic capacitance $C\textsubscript{p}$ at approximately \qty{100}{\pF}, based on the observed shift in $f_{\text{lo},1}$ for small values of $C_1$ (e.g.\ \qty{150}{\pF}). Here, $C\textsubscript{p}$ is an effective parallel feedthrough across $C_1$ in the VNA setup (coupler/launch geometry and fixture parasitics), not the CPWG or connector capacitance alone. For \(C_1\gg C\textsubscript{p}\) the influence of parasitic capacitances becomes negligible. As noted earlier, the two-port VNA configuration shares a common instrument ground at both ports; consequently, $C_2$ does not carry signal current in these measurements and the VNA-derived cutoff frequencies constrain $C_1$ and its effective parallel parasitics $C\textsubscript{p}$.
\begin{equation}
    f_{\text{lo,1}} = \frac{1}{4\pi Z_0 C_{\text{eff},1}} = \frac{1}{4\pi Z_0 (C_1 + C\textsubscript{p})}
    \label{eq:tau1}
\end{equation}

\begin{figure*}
    \centering
    \includegraphics[width=.99\textwidth]{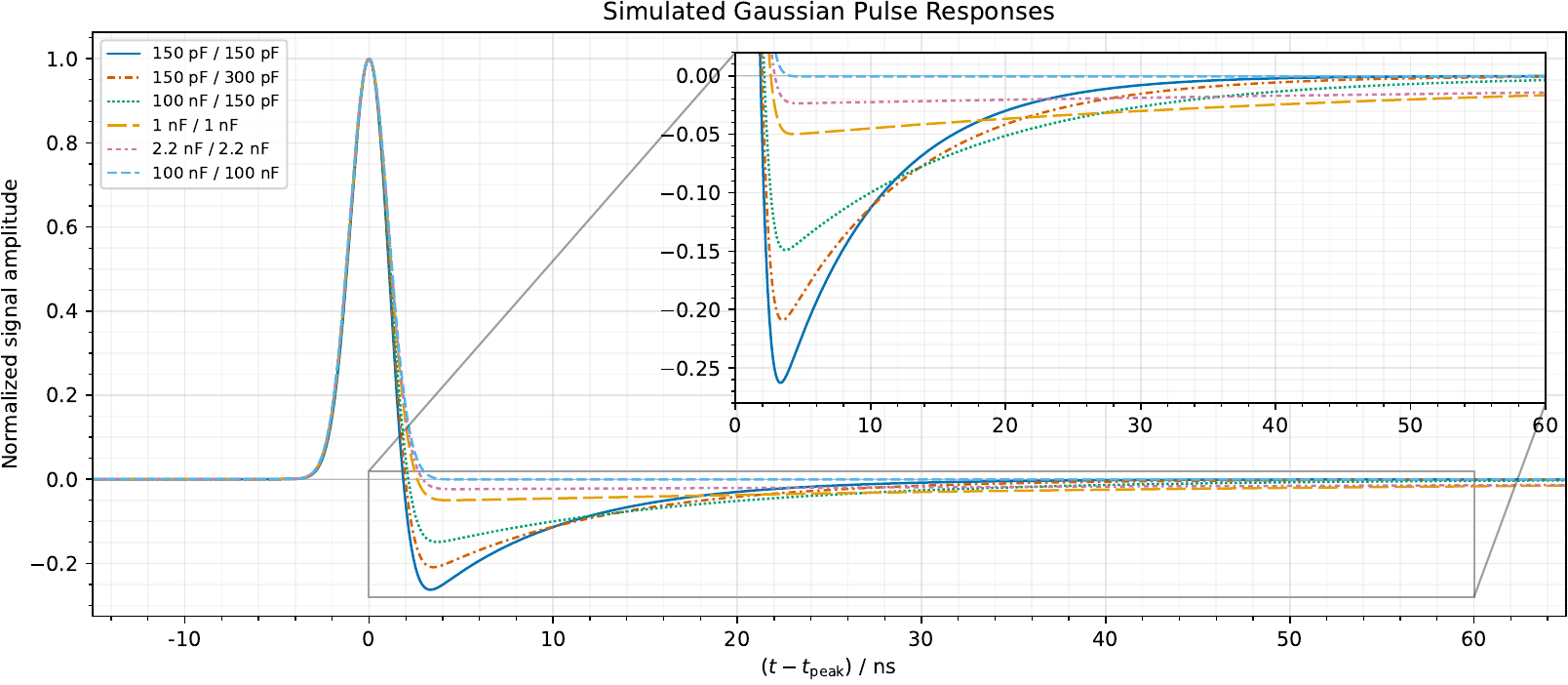}
    \caption{\label{fig:simulink} Time-domain Gaussian pulse responses obtained from the Simulink circuit-envelope transient simulation for different capacitor values; legend entries denote the capacitor pair $(C_1\; /\; C_2)$ used in the simulation.}
\end{figure*}

The circuit-level transient model provides additional insight into how these high-pass cutoff frequencies manifest themselves in the time domain. In particular, the undershoot-recovery time constants $\tau$ are accurately reproduced by both presented modeling approaches, reflecting the combined contributions of $C_1$ and $C_2$ as given by Eq.~(\ref{eq:tau12}) and ultimately based on Eq.~(\ref{eq:rc}) (see Appendix~\ref{sec:abcd} for the derivation). This analysis leads to the cutoff frequency values $f_{\text{lo},2}$. The Gaussian impulse responses from the Simulink model for various component values are shown in Fig.~\ref{fig:simulink}.

Equation~(\ref{eq:tau12}) is a lumped-element expression and is used here specifically to describe the low-frequency high-pass corner. It is valid as long as $f_{\text{lo},2}$ lies well below the self-resonant frequencies of $C_1$ and $C_2$, such that both behave capacitively in this range. For strongly asymmetric values (e.g.\ iteration~C with $C_2\gg C_1$), \(C_{\text{eff},2} \approx C_1\), the cutoff is essentially insensitive to the frequency dependence of $C_2$ in the vicinity of $f_{\text{lo},2}$.
\begin{equation}
    f_{\text{lo,2}} = \frac{1}{4\pi Z_0 C_{\text{eff},2}} = \frac{1}{4\pi Z_0\frac{C_1 C_2}{C_1+C_2}}
    \label{eq:tau12}
\end{equation}

\begin{table*}
    \caption{\label{tab:vna_freqs} Detector capacitor values and derived cutoff frequencies via VNA measurements and circuit modeling. The high-pass corners $f_{\text{lo},1}$ and $f_{\text{lo},2}$ follow from the coupling network. The reported $f\textsubscript{hi}$ is the low-pass corner of the assembled signal path which includes the anode response as well as implementation- and fixture-dependent high-frequency effects.}
    \begin{ruledtabular}
    \begin{tabular}{lrrrrrr}
            Iteration & $C_1$ & $C_2$ & $C_{\text{eff},2}$ & $f_{\text{lo},1}$ & $f_{\text{lo},2}$ & $f\textsubscript{hi}$ \\
        \hline
            A & \qty{150}{\pF} & \qty{300}{\pF} & \qty{100}{\pF} & \qty{5.63}{\MHz} & \qty{15.9}{\MHz} & \qty{2.12}{\GHz} \\
            B & \qty{1}{\nF} & \qty{1}{\nF} & \qty{0.5}{\nF} & \qty{1.56}{\MHz} & \qty{3.18}{\MHz} & \qty{1.63}{\GHz} \\
            C & \qty{100}{\nF} & \qty{150}{\pF} & \qty{150}{\pF} & \qty{14.3}{\kHz} & \qty{10.6}{\MHz} & \qty{2.08}{\GHz} \\
            D & \qty{100}{\nF} & \qty{100}{\nF} & \qty{50}{\nF} & \qty{14.4}{\kHz} & \qty{31.8}{\kHz} & \qty{1.91}{\GHz} \\
            E & \qty{2.2}{\nF} & \qty{2.2}{\nF} & \qty{1.1}{\nF} & \qty{694}{\kHz} & \qty{1.45}{\MHz} & \qty{2.13}{\GHz} \\
    \end{tabular}
    \end{ruledtabular}
\end{table*}

\subsection{Pulse-shape improvements}

\begin{figure*}
    \centering
    \includegraphics[width=.99\textwidth]{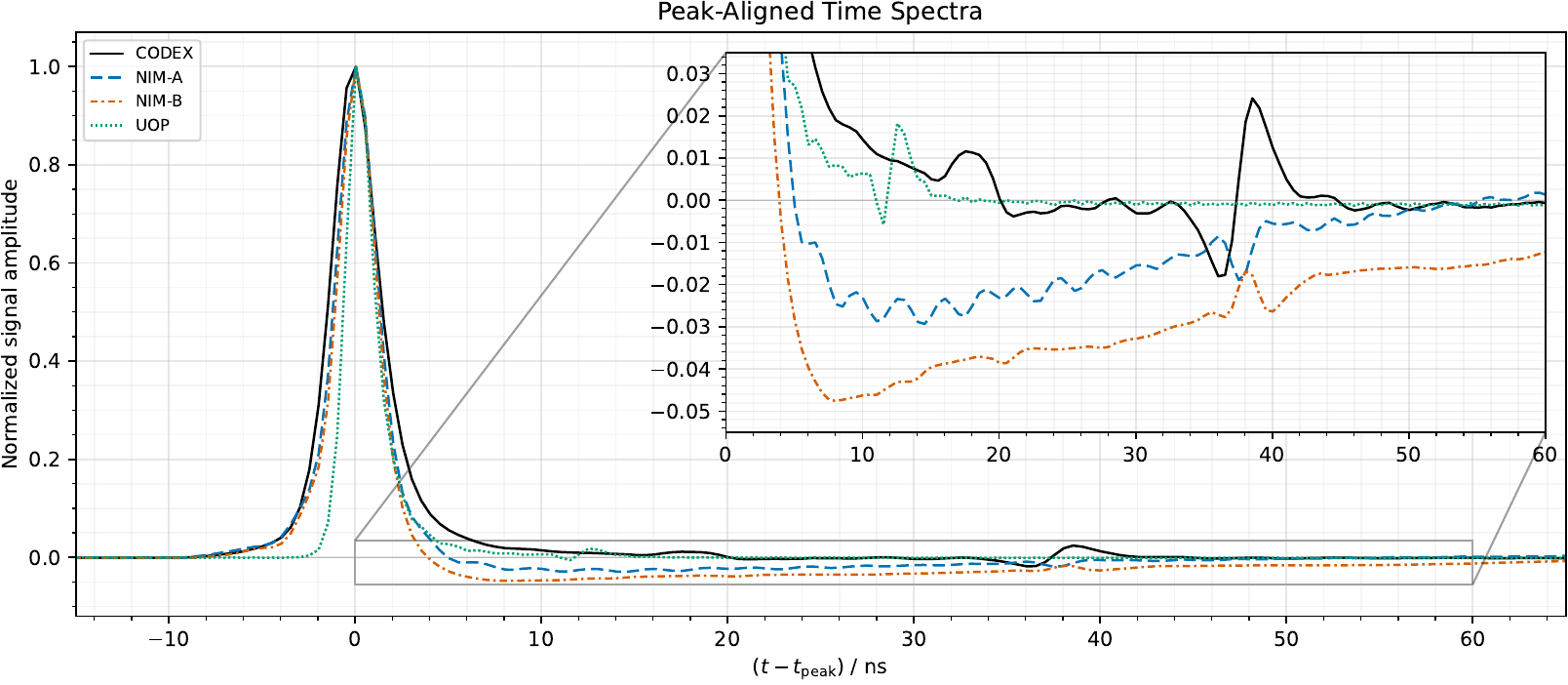}
    \caption{\label{fig:results_compare} Measured peak-aligned, amplitude-normalized end-to-end time responses under ion-saturation operation for different detector architectures. Traces correspond to the highest-amplitude low-mass peak (typically \ce{H2O} after mass calibration). CODEX, NIM-A, and NIM-B were acquired in the MEFISTO test setup while UOP was acquired in a separate setup.}
\end{figure*}

End-to-end mass spectrometric measurements confirm the trends observed in the isolated simulation and testing steps. Figure~\ref{fig:results_compare} compares the detector's time response under saturation conditions with the established reference architectures. The plotted waveforms correspond to the highest-amplitude low-mass peak in each spectrum (typically \ce{H2O} after mass calibration) and are peak-aligned and normalized in amplitude to enable a shape-based comparison; polarity is shown positive-going by convention.

This comparison is therefore intended as a shape-based end-to-end benchmark rather than as a strict detector-only comparison under identical MCP, bias, and decoupling conditions. CODEX, NIM-A, and NIM-B were measured in the same MEFISTO test setup, whereas UOP was acquired in a separate setup; therefore, absolute amplitudes are not compared.

To enable a quantitative assessment, the time-domain metrics introduced earlier were applied to all spectra, with the resulting values summarized in Table~\ref{tab:metrics}.

\begin{table*}
    \caption{\label{tab:metrics} Pulse-shape metrics per detector extracted from the measured, peak-aligned and amplitude-normalized waveforms shown in Fig.~\ref{fig:results_compare}.}
    \begin{ruledtabular}
    \begin{tabular}{lrrrr}
        Metric & \text{CODEX} & \text{NIM-A} & \text{NIM-B} & \text{UOP} \\
        \hline
        Transition duration (positive-going) & \qty{2.39}{\ns} & \qty{2.56}{\ns} & \qty{2.49}{\ns} & \qty{1.27}{\ns} \\
        Transition duration (negative-going) & \qty{3.38}{\ns} & \qty{2.34}{\ns} & \qty{2.06}{\ns} & \qty{2.34}{\ns} \\
        Pulse duration & \qty{2.98}{\ns} & \qty{2.57}{\ns} & \qty{2.38}{\ns} & \qty{1.80}{\ns} \\
        Asymmetry index & \qty{-17.2}{\percent} & \qty{4.4}{\percent} & \qty{9.6}{\percent} & \qty{-29.6}{\percent} \\
        Undershoot & \qty{< 0.1}{\percent} & \qty{2.5}{\percent} & \qty{4.6}{\percent} & \qty{< 0.1}{\percent} \\
        Tail time constant $\tau$ & \qty{4.44}{\us} & \qty{45.5}{\ns} & \qty{48.3}{\ns} & \qty{2.26}{\us} \\
        Settling duration (\qty{1.0}{\percent}) & \qty{39.1}{\ns} & \qty{37.7}{\ns} & \qty{170.8}{\ns} & \qty{12.4}{\ns} \\
        Settling duration (\qty{0.1}{\percent}) & \qty{4.53}{\us} & \qty{2.48}{\us} & \qty{2.39}{\us} & \qty{3.97}{\us} \\
    \end{tabular}
    \end{ruledtabular}
\end{table*}
Qualitative and quantitative comparison of measured time series and derived metrics revealed three key insights detailed in the following.

First, the proposed CODEX detector achieves signal fidelity on par with the UOP waveguide-based detector by \citeauthor{wurz_fast_1996}, \cite{wurz_fast_1996} demonstrating that a planar implementation can match the timing quality of the waveguide architecture while using substantially less volume and mass (reducing the detector length along the ion-flight direction by approximately a factor of three and lowering the detector mass by nearly an order of magnitude, excluding the casing). The CODEX detector also offers increased mechanical robustness, withstanding mechanical stresses of up to \qty{40}{g_{rms}}. \cite{Fausch2026} Although the \qty{1}{\percent} settling time is slightly longer than for UOP, this difference is attributable to small post-peak reflections in the test setup, which marginally bias the metric. Compact MCP detectors have been demonstrated previously, such as in suspended-substrate configurations, \cite{schletti_fast_2001} nevertheless, the presented design further reduces complexity by integrating the entire detector into a single planar PCB without suspended-substrate structures or other elements that require mechanical extension orthogonal to the anode surface.

Second, the detectors exhibit varying degrees of post-peak undershoot, with the CODEX and UOP designs showing the smallest amplitudes. The spiral-based and patch-based NIM prototype detectors display undershoot amplitudes at the percent level, with the patch variant (NIM-A) performing slightly better than the spiral (NIM-B), highlighting the influence of anode geometry on the transient response. In contrast, the CODEX and UOP detectors show undershoot amplitudes below \qty{1}{\percent} and only minor baseline distortions at the \num{e-3} level (i.e.\ $<\qty{0.1}{\percent}$ of the peak amplitude).

Third, the UOP detector exhibits a microsecond-scale exponential recovery tail, reflected by its fitted tail time constant. This behavior is consistent with the larger effective anode-to-ground capacitance expected for bulkier waveguide anode implementations, which can dominate the slow baseline recovery in the absence of pronounced undershoot after the main peak. A similar dependence of the post-peak tail on anode capacitance has been reported previously.\cite{kuzminchuk_performance_2011} In the present comparison, both CODEX and UOP exhibit undershoot amplitudes below \qty{0.1}{\percent}; consequently, the fitted $\tau$ values describe the slow residual baseline re-centering at the \num{e-3} level rather than a large-amplitude tail. Therefore, the improvement of CODEX relative to the NIM prototype detectors is the suppression of percent-level post-peak undershoot to below \qty{0.1}{\percent}, whereas the comparison with UOP demonstrates that CODEX achieves similarly low undershoot and comparable timing fidelity in a substantially more compact planar implementation.

All detector designs tested in the MEFISTO setup (this includes CODEX, NIM-A, and NIM-B) show reflective peaks around \qty{40}{\ns} post-peak. For CODEX, the relative amplitude of these peaks is higher than for the NIM detectors because of the absence of source termination. Nevertheless, these effects can be traced unambiguously back to imperfect terminations of the readout system of the test setup and can therefore be decoupled from the performance of the detector itself. The UOP waveform was acquired in a separate setup and, therefore, does not share MEFISTO-specific artifacts.

\section{Discussion}

\subsection{Current return path}

When comparing the pulse response derived from the VNA measurements with the spectra obtained from end-to-end testing, we find that the influence of the $RC$ components is less pronounced in the physical system than in the idealized circuit model. This discrepancy arises because, in integrated TOF-MS operation, the detector current return path is not limited to the two transmission-line legs containing $C_1$ and $C_2$. The MCP stack and its biasing network introduce additional return paths and stray capacitances, which alter the effective time constants observed in full mass spectrometric measurements. As a result, the simple balanced-anode assumption becomes insufficient to capture the full system behavior.

When the MCP stack is included in the model, the excitation must be treated as a differentially propagating mode between the MCP and anode potentials. In this configuration, the return current can flow through the bias network toward $U_1$ and $U_2$ only insofar as the network presents a sufficiently low impedance at the signal frequencies. Practically, this requires the bias network to offer a return path with an effective time constant $\tau\textsubscript{eff}$ shorter than the signal reference on the detector. Cable and feedthrough stray capacitances may contribute, but are not well controlled and therefore are treated as incidental rather than a design element.

An additional capacitor between the signal line and the MCP could be introduced to further stabilize the floating segment potential during fast transients. \cite{schletti_fast_2001} However, our measurements have shown that acceptable signal integrity is already achieved without dedicated lumped capacitance.

To verify that the observed return paths do not compromise MCP gain stability in ion-saturation operation, we next estimate the charge budget and compare the implied average output current to the MCP strip current.

\subsection{MCP rate handling}

In standard operation, our MCP is biased to a nominal analog-mode gain of \(G\approx\num{e6}\), which we use here as an order-of-magnitude estimate for the transient charge budget, such that a single detected ion generates an anode charge pulse of \(Q_1=Ge\approx\qty{1.6e-13}{\C}\). For representative low-mass peaks in ion saturation mode, we infer on the order of \(N\approx100\) ions within a pulse of \(t\textsubscript{FWHM}\approx\qty{1.5}{\ns}\), corresponding to a total charge of \(Q\approx NQ_1 \approx \qty{1.6e-11}{\C}\) and a peak current of \(I\textsubscript{max} \approx Q/t\textsubscript{FWHM} \approx \qty{10}{\mA}\). That is, a pulse amplitude of about \qty{0.5}{\V} on a \qty{50}{\ohm} load. Importantly, $I\textsubscript{max}$ is the transient current delivered into the readout. If we now conservatively assume that this charge is drawn in first order from the local capacitances supporting the floating segment, \(C\textsubscript{p}\approx\qty{100}{\pF}\), the associated transient voltage drop at the floating node is only \(\Delta V = Q/C\textsubscript{p} \approx \qty{0.16}{\V}\). We therefore consider the voltage drop induced by individual pulses to be negligible for the MCP gain and timing performance. For larger saturated peaks distributed over several nanoseconds, the total number of ions may be substantially higher; in that regime, the relevant limitation shifts from the instantaneous transient to the average extracted charge per spectrum.

On longer timescales, however, MCP rate limitations are governed by the average output current per spectrum relative to the MCP strip current \(I\textsubscript{strip} = U\textsubscript{MCP}/R\textsubscript{MCP}\) rather than by single-event dynamics. Typical values for $R\textsubscript{MCP}$ are on the order of \num{e8} to \qty{e9}{\ohm}. \cite{wiza_microchannel_1979} For our setup, we use  \(U\textsubscript{MCP} = \qty{2}{\kV}\), \(R\textsubscript{MCP}=\qty{2e8}{\ohm}\) and pulse our ion extraction at a frequency of \(f\textsubscript{pulse} = \qty{10}{\kHz}\). The time per individual spectrum therefore corresponds to \(t\textsubscript{spectrum} = 1/f\textsubscript{pulse} = \qty{100}{\us}\).

Assuming $N\textsubscript{peak}=5$ saturated peaks per spectrum, the average output current can be estimated as \(I\textsubscript{avg} = QN\textsubscript{peak} / t\textsubscript{spectrum} \) and compared with the MCP strip current, as shown in  Eq.~(\ref{eq:mcp_limit}). For the limit value of \(0.1\cdot I\textsubscript{strip}\), we can assume constant gain. \cite{wiza_microchannel_1979}
\begin{multline}
    \frac{QN\textsubscript{peak}}{t\textsubscript{spectrum}} = I\textsubscript{avg} \approx \qty{0.8}{\uA} \\
    < \qty{1.0}{\uA} \approx 0.1 \cdot I\textsubscript{strip} = 0.1 \cdot \frac{U\textsubscript{MCP}}{R\textsubscript{MCP}}
    \label{eq:mcp_limit}
\end{multline}

\subsection{Implication of high-pass behavior}

The lower cutoff frequency of the detector determines how the baseline recovers after a pulse. Different return paths contribute to this recovery, each associated with their own characteristic time constants. The voltage drop on the MCP is rapidly restored through local capacitances and does not lead to pronounced undershoot. In contrast, the redistribution of charge needed to re-establish the anode baseline occurs through much larger resistive paths, analogous to the slow recharge observed in photomultiplier tubes. \cite{genolini_design_2001} These slow paths introduce only a small baseline offset, but ensure a nearly linear post-pulse baseline. We optimize our design for a long recovery time to maintain a nearly linear (although slightly offset) baseline signal, which simplifies the interpretation of the obtained mass spectrum, especially in the presence of multiple adjacent peaks.

From a frequency-domain perspective, AC coupling suppresses the DC component and attenuates the low-frequency part of the pulse spectrum. As a result, the coupled waveform must recover toward a zero-mean baseline: A positive detector pulse therefore acquires a compensating post-peak undershoot. The magnitude of the undershoot increases as the high-pass corner frequency $f_c$ approaches the characteristic spectral content of the pulse, because a larger fraction of low-frequency content is removed. In contrast, choosing $f_c \ll 1/t\textsubscript{FWHM}$ keeps the main lobe mostly unchanged and affects mainly the recovery after the pulse, at the cost of a longer recovery time constant $\tau\textsubscript{eff} = 1/(2\pi f_c)$.

In addition, for the CODEX implementation, the dominant slow equalization path on the floating detector side is provided by the large bleed resistor $R_3$, which defines a DC return between the two signal legs without appreciably shunting the fast transient. This path primarily controls slow baseline re-centering and baseline offset evolution between events.

\subsection{Impedance}

The detector signal path is modeled as a transmission line consisting of a localized anode source impedance, followed by a \qty{50}{\ohm} coplanar waveguide across the decoupling capacitor toward the SMA interface, leading to a coaxial cable and eventually the readout electronics. The impedance of the via connecting the anode to the CPWG on the opposing PCB side is ignored in this simplified argument, as its length is much shorter than the critical electrical length at the relevant frequencies, as derived in Eq.~(\ref{eq:lambda}).

The source impedance of the anode assumes a first minimum at the zeroth-order resonance. As our design operates far below the resonance, the anode input impedance is dominated by its capacitance to the local return (bottom-layer ground and any nearby shielding), i.e.\ \(Z\textsubscript{s}(\omega) \approx 1/(j\omega C_{\text{a,GND}})\) rather than a purely resistive open circuit. For a patch with a radius of a few \unit{\mm} on FR-4 substrate, $C_{\text{a,GND}}$ is typically on the order of a few picofarads by a parallel-plate estimate, implying \(\lvert Z\textsubscript{s}\rvert \approx\qty{e3}{\ohm}\) at \qty{100}{\MHz} and \(\lvert Z\textsubscript{s}\rvert \approx\qty{e2}{\ohm}\) at \qty{1}{\GHz}.

Consequently, the anode appears approximately open (\(\Gamma\approx +1\)) only over the frequency range where \(\lvert Z\textsubscript{s}(\omega) \rvert \gg Z_0\) as shown in Eq.~(\ref{eq:refl}). For the local anode capacitance on the order of a few \unit{\pF}, this boundary lies in the few-hundred-\unit{\MHz} to low-\unit{\GHz} regime and therefore overlaps with the dominant low-frequency content of Gaussian detector pulses. Maintaining signal integrity therefore requires that downstream reflections which could travel back to the anode be minimized through careful impedance matching across all intermediate connectors and interfaces toward the readout system.
\begin{equation}
    \Gamma = \frac{Z\textsubscript{s} - Z_0}{Z\textsubscript{s} + Z_0} \xrightarrow{\lvert Z\textsubscript{s}\rvert / Z_0 \,\to\,\infty} +1
    \label{eq:refl}
\end{equation}

Introducing a source termination (e.g., a shunt resistor $R\textsubscript{s}$ at the anode) would reduce $\lvert\Gamma\rvert$ but at the cost of peak amplitude and signal-to-noise ratio: A current source drive into $Z_0\parallel R\textsubscript{s}$ diverts part of the signal current into $R\textsubscript{s}$ and adds thermal noise. As the goal of our design is to enable mass spectrometric measurements which are as sensitive as possible, we therefore omit source termination and instead maximize current transfer from the anode into the transmission line and subsequent readout. An active buffer (wideband amplifier) placed immediately after the anode and AC-decoupling stage could, in principle, suppress backward-propagating reflections via its reverse isolation. We did not pursue this approach here because it conflicts with the floating high-voltage architecture and spaceborne constraints such as power and qualification; it remains a potential mitigation option for future iterations.

Nevertheless, matched anodes with controlled impedance transitions were previously derived and realized for fast-timing applications by \citeauthor{wurz_impedancematching_1994}, \cite{wurz_impedancematching_1994} leading to the UOP detector design used here as a reference. \cite{wurz_fast_1996} Furthermore, nearly planar designs have also been demonstrated in suspended-substrate configurations. \cite{schletti_fast_2001}

In our measurements, the proposed CODEX design exhibits the strongest apparent reflections among the detectors tested. However, these reflections originate from imperfect connector and cable terminations as part of the MEFISTO test setup. Due to the absence of source termination, these issues are exacerbated with respect to the other detectors studied. In a final implementation, particularly in a flight instrument, where the requirement of strict impedance matching with high-quality connectors and cables will be met, the residual reflections attributable to the detector are expected to be negligible.

\subsection{Final design}

For the final developed detector architecture, we converged on a planar patch anode with anode-proximal AC decoupling and no resistive source termination. Specifically, the distance from the anode to the decoupling capacitors is chosen to satisfy the condition derived in Eq.~(\ref{eq:lambda}). The patch anode confines the electric field and avoids the distributed mode delay spread observed with spiral geometries. The detector layout employs CPWG structures for the signal line with a characteristic impedance of \qty{50}{\ohm}. Decoupling capacitors $C_1$ and $C_2$ were placed on opposing PCB layers to provide symmetric HV confinement for detector bias voltages up to \qty{-2}{\kV}.

Furthermore, $C_1$ and $C_2$ are RF-grade Class-I C0G capacitors selected with self-resonant frequencies well above the detector pulse bandwidth; Class-II X7R parts are avoided due to bias-dependent capacitance and higher loss at elevated frequency.

The component values used in the final design are listed in Table~\ref{tab:vals} and were selected to achieve the desired undershoot amplitude and baseline-recovery characteristics. Using these component values, the effective series coupling capacitance is $C_{\text{eff},2}=\qty{0.6}{\nF}$, which corresponds to an effective high-pass time constant of $\tau\textsubscript{eff}\approx \qty{60}{\ns}$ under \qty{50}{\ohm}-terminated readout conditions (\(R\textsubscript{eff} = 2Z_0\)). The resistor $R_1$ is omitted, as the MCP stack already provides a sufficiently high-impedance path and does not require an additional bias current. $R_2$ was substituted by a Zener diode, maintaining a constant voltage bias between the MCP backside and the anode to sufficiently accelerate the electrons released by the MCP to the anode to avoid a transient time spread. The resistor $R_3$ that connects the two signal legs must remain large to avoid shunting transient currents away from the \qty{50}{\ohm} transmission line, consistent with the rationale for omitting source termination. In principle, $R_3$ could be implemented as an RF choke (bias-tee concept) to provide a low-impedance DC equalization path while maintaining high impedance over the detector pulse bandwidth. This option was not explored in the present implementation and is left for future iterations, as practical inductors exhibit parasitics and a finite self-resonant frequency that can introduce additional resonances at high frequency.

\begin{table}
\centering
    \caption{\label{tab:vals} Final detector component values used in the CODEX implementation (definitions as in Fig.~\ref{fig:final_circ}). $C_1$/$C_2$ are the AC-decoupling capacitors, $R_2$ fixes the anode bias, $R_3$ is a high-value bleed resistor, and $R_4$ is a small series damping element.}
\begin{minipage}{0.5\linewidth}
    \begin{ruledtabular}
    \begin{tabular}{lr}
        $C_1$ & \qty{1.2}{\nF} \\
        $C_2$ & \qty{1.2}{\nF} \\
        $R_1$ & Open \\
        $R_2$ & \qty{180}{\V} Zener diode \\
        $R_3$ & \qty{100}{\kilo\ohm} \\
        $R_4$ & \qty{20}{\ohm} \\
    \end{tabular}
    \end{ruledtabular}
\end{minipage}
\end{table}

\section{Conclusion}

We presented a time-domain co-design method and architecture for a floating-anode TOF-MS MCP detector that integrates the collector surface and the high-voltage AC-coupling stage into a unified design. The developed detector meets the single-event timing requirements of compact TOF-MS instruments and substantially suppresses post-peak artifacts, such as undershoot amplitude and long settling-time tails, without sacrificing pulse amplitude. We validated the design through a staged workflow that included partial and full simulations together with end-to-end mass spectrometric measurements, evaluated with IEEE-consistent metrics. We further linked the low-frequency cutoff set by the decoupling network to the time-domain response, including the undershoot decay behavior.

Together, these results yield a set of practical design rules for compact floating-anode detectors: place and size the decoupling capacitors to push the cutoff frequency below the analysis window, minimize the high-frequency return loop, smooth the few unavoidable impedance transitions, and reserve resistive termination only when downstream impedance matching is not applicable.

Our resulting planar circular patch anode achieves waveguide-level pulse fidelity at a fraction of the mass and volume of waveguide-based designs, while improving mechanical robustness and offering a clear scaling path to larger active areas for future instruments. This provides a flight-ready, manufacturable, and electrically stable architecture for next-generation spaceborne TOF-MS detectors.

Variants of this architecture are being adopted in several next-generation TOF-MS developments, including CODEX, \cite{Fausch2026, Wurz2026} CubeSatTOF, \cite{fausch_direct_2022} OpenTOF, \cite{schertenleib_ion-optical_2024} and the Neutral Gas Mass Spectrometer (NGMS). \cite{fausch_advancing2_2025} This demonstrates that the architecture is not only technically viable but is already being integrated into the development of multiple instruments.

\begin{acknowledgments}
    This work was supported by the Swiss National Science Foundation (SNSF) [207312] and the Canton of Bern. For the purpose of open access, a CC BY public copyright license is applied to any author-accepted manuscript (AAM) version arising from this submission. Additional support from the Swiss Space Office (SSO) is gratefully acknowledged.
    
    We thank the reviewers for their careful reading of the manuscript and their constructive comments, which helped improve the clarity and quality of this work.
\end{acknowledgments}

\section*{Conflict of interest statement}
The authors have no conflicts of interest to disclose.
    
\section*{Data availability statement}
The data that support the findings of this study are available from the corresponding author upon reasonable request.

\appendix

\section{Loaded Two-Port Model}
\label{sec:abcd}

We model the decoupling network as a uniform transmission line section loaded by a series impedance, representing the capacitive loading due to the AC-coupling element. The cutoff frequency can be derived from the $ABCD$-parameters of a series impedance $Z(\omega)$ between two ports:
\begin{equation}
  \begin{bmatrix}
    A & B \\ C & D
  \end{bmatrix}
  =
  \begin{bmatrix}
    1 & Z(\omega) \\
    0 & 1
  \end{bmatrix}.
\end{equation}

The transmission coefficient $S_{21}(\omega)$ for a reciprocal passive two-port with reference impedance $Z_0$ is given by
\begin{equation}
  S_{21}(\omega) = \frac{2}{A + B/Z_0 + C Z_0 + D}.
\end{equation}
Substitution leads to the following expression:
\begin{equation}
  S_{21}(\omega) = \frac{2}{2 + Z(\omega)/Z_0}.
\end{equation}
In our generalized case, the impedance consists of a capacitor and we define an equivalent capacitance $C\textsubscript{eq}$ such that
\begin{equation}
  Z(\omega) = \frac{1}{j\omega C_{\text{eq}}}.
\end{equation}
Inserting this expression into $S_{21}$ yields
\begin{equation}
  S_{21}(\omega)
  = \frac{2}{2 - \dfrac{j}{\omega Z_0 C_{\text{eq}}}},
\end{equation}
with magnitude
\begin{equation}
  \lvert S_{21}(\omega) \rvert
  = \frac{2}{\sqrt{4 +
  \left( \dfrac{1}{\omega Z_0 C_{\text{eq}}} \right)^2 }}.
\end{equation}
The low-frequency cutoff point $f_{\text{lo}}$ is defined at the \qty{-3}{\dB} point $\lvert S_{21}(\omega_c) \rvert = 1/\sqrt{2}$, which gives
\begin{equation}
  \omega_c = \frac{1}{2 Z_0 C_{\text{eq}}}
  \quad\Rightarrow\quad
  f_{\text{lo}} = \frac{\omega_c}{2\pi}
  = \frac{1}{4\pi Z_0 C_{\text{eq}}}.
\end{equation}

In the main text, we use the equivalent capacitances $C_{\text{eff},1}$ and $C_{\text{eff},2}$ for the single- and dual-capacitor configurations, leading to Eqs.~(\ref{eq:tau1}) and (\ref{eq:tau12}):
\begin{equation}
  C_{\text{eff},1} = C_1 + C\textsubscript{p}
  ,\qquad
  C_{\text{eff},2} = \frac{C_1 C_2}{C_1 + C_2}.
\end{equation}

\bibliography{references}

@inproceedings{fohn_description_2021,
    title = {Description of the {Mass} {Spectrometer} for the {Jupiter} {Icy} {Moons} {Explorer} {Mission}},
    url = {https://ieeexplore.ieee.org/document/9438344},
    doi = {10.1109/AERO50100.2021.9438344},
    urldate = {2024-10-02},
    booktitle = {2021 {IEEE} {Aerospace} {Conference}},
    author = {Föhn, Martina and Galli, André and Vorburger, Audrey and Tulej, Marek and Lasi, Davide and Riedo, Andreas and Fausch, Rico G. and Althaus, Michael and Brüngger, Stefan and Fahrer, Philipp and Gerber, Michael and Lüthi, Matthias and Munz, Hans Peter and Oeschger, Severin and Piazza, Daniele and Wurz, Peter},
    month = mar,
    year = {2021},
    pages = {1--14},
}

@article{meyer_fully_2017,
    title = {Fully automatic and precise data analysis developed for time-of-flight mass spectrometry},
    volume = {52},
    copyright = {Copyright © 2017 John Wiley \& Sons, Ltd.},
    issn = {1096-9888},
    url = {https://onlinelibrary.wiley.com/doi/abs/10.1002/jms.3964},
    doi = {10.1002/jms.3964},
    language = {en},
    number = {9},
    urldate = {2025-10-08},
    journal = {Journal of Mass Spectrometry},
    author = {Meyer, Stefan and Riedo, Andreas and Neuland, Maike B. and Tulej, Marek and Wurz, Peter},
    year = {2017},
    pages = {580--590},
}

@inproceedings{fausch_direct_2022,
    title = {Direct {Measurement} of {Neutral} {Gas} during {Hypervelocity} {Planetary} {Flybys}},
    url = {https://ieeexplore.ieee.org/document/9843767},
    doi = {10.1109/AERO53065.2022.9843767},
    urldate = {2025-10-02},
    booktitle = {2022 {IEEE} {Aerospace} {Conference}},
    author = {Fausch, Rico G. and Wurz, Peter and Cotting, Björn and Rohner, Urs and Tulej, Marek},
    month = mar,
    year = {2022},
    pages = {1--12},
}

@article{levine_dating_2023,
    title = {Dating {Granites} {Using} {CODEX}, with {Application} to {In} {Situ} {Dating} on the {Moon}},
    volume = {4},
    issn = {2632-3338},
    url = {https://iopscience.iop.org/article/10.3847/PSJ/accd6c/meta},
    doi = {10.3847/PSJ/accd6c},
    language = {en},
    number = {5},
    urldate = {2025-10-08},
    journal = {The Planetary Science Journal},
    author = {Levine, Jonathan and Anderson, F. Scott and Braden, Sarah and Fausch, Rico G. and Foster, Sean and Fowler, Gavin and Joy, Katherine H. and Osterman, Steven and Pernet-Fisher, John and Seddio, Stephen and Whitaker, Tom and Wurz, Peter and Yant, Marcella and Yap, Teng Ee},
    month = may,
    year = {2023},
    pages = {92},
}

@article{arevalo_jr_mass_2020,
    title = {Mass spectrometry and planetary exploration: {A} brief review and future projection},
    volume = {55},
    issn = {1096-9888},
    shorttitle = {Mass spectrometry and planetary exploration},
    url = {https://onlinelibrary.wiley.com/doi/abs/10.1002/jms.4454},
    doi = {10.1002/jms.4454},
    language = {en},
    number = {1},
    urldate = {2024-08-08},
    journal = {Journal of Mass Spectrometry},
    author = {Arevalo Jr, Ricardo and Ni, Ziqin and Danell, Ryan M.},
    year = {2020},
    pages = {e4454},
}

@article{wiza_microchannel_1979,
    title = {Microchannel plate detectors},
    volume = {162},
    issn = {0029-554X},
    url = {https://www.sciencedirect.com/science/article/pii/0029554X79907341},
    doi = {10.1016/0029-554X(79)90734-1},
    number = {1},
    urldate = {2024-10-02},
    journal = {Nuclear Instruments and Methods},
    author = {Wiza, Joseph Ladislas},
    month = jun,
    year = {1979},
    pages = {587--601},
}

@inproceedings{lasi_decisions_2020,
    title = {Decisions and {Trade}-{Offs} in the {Design} of a {Mass} {Spectrometer} for {Jupiter}'s {Icy} {Moons}},
    url = {https://ieeexplore.ieee.org/document/9172784},
    doi = {10.1109/AERO47225.2020.9172784},
    urldate = {2024-10-02},
    booktitle = {2020 {IEEE} {Aerospace} {Conference}},
    author = {Lasi, Davide and Meyer, Stefan and Piazza, Daniele and Lüthi, Mathias and Nentwig, Andreas and Gruber, Mario and Brüngger, Stefan and Gerber, Michael and Braccini, Saverio and Tulej, Marek and Föhn, Martina and Wurz, Peter},
    month = mar,
    year = {2020},
    pages = {1--20},
}

@article{riedo_high-speed_2017,
    title = {High-speed microstrip multi-anode multichannel plate detector system},
    volume = {88},
    issn = {0034-6748},
    url = {https://doi.org/10.1063/1.4981813},
    doi = {10.1063/1.4981813},
    number = {4},
    urldate = {2024-10-03},
    journal = {Review of Scientific Instruments},
    author = {Riedo, Andreas and Tulej, Marek and Rohner, Urs and Wurz, Peter},
    month = apr,
    year = {2017},
    pages = {045114},
}

@article{abplanalp_neutral_2009,
    title = {A neutral gas mass spectrometer to measure the chemical composition of the stratosphere},
    volume = {44},
    issn = {0273-1177},
    url = {https://www.sciencedirect.com/science/article/pii/S0273117709004360},
    doi = {10.1016/j.asr.2009.06.016},
    number = {7},
    urldate = {2025-10-01},
    journal = {Advances in Space Research},
    author = {Abplanalp, D. and Wurz, P. and Huber, L. and Leya, I. and Kopp, E. and Rohner, U. and Wieser, M. and Kalla, L. and Barabash, S.},
    month = oct,
    year = {2009},
    pages = {870--878},
}

@article{waite_maspex-europa_2024,
    title = {{MASPEX}-{Europa}: {The} {Europa} {Clipper} {Neutral} {Gas} {Mass} {Spectrometer} {Investigation}},
    volume = {220},
    issn = {1572-9672},
    shorttitle = {{MASPEX}-{Europa}},
    url = {https://doi.org/10.1007/s11214-024-01061-6},
    doi = {10.1007/s11214-024-01061-6},
    language = {en},
    number = {3},
    urldate = {2024-10-02},
    journal = {Space Science Reviews},
    author = {Waite, J. H. and Burch, J. L. and Brockwell, T. G. and Young, D. T. and Miller, G. P. and Persyn, S. C. and Stone, J. M. and Wilson, P. and Miller, K. E. and Glein, C. R. and Perryman, R. S. and McGrath, M. A. and Bolton, S. J. and McKinnon, W. B. and Mousis, O. and Sephton, M. A. and Shock, E. L. and Choukroun, M. and Teolis, B. D. and Wyrick, D. Y. and Zolotov, M. Y. and Ray, C. and Magoncelli, A. L. and Raffanti, R. R. and Thorpe, R. L. and Bouquet, A. and Salter, T. L. and Robinson, K. J. and Urdiales, C. and Tyler, Y. D. and Dirks, G. J. and Beebe, C. R. and Fugett, D. A. and Alexander, J. A. and Hanley, J. J. and Moorhead-Rosenberg, Z. A. and Franke, K. A. and Pickens, K. S. and Focia, R. J. and Magee, B. A. and Hoeper, P. J. and Aaron, D. P. and Thompson, S. L. and Persson, K. B. and Blase, R. C. and Dunn, G. F. and Killough, R. L. and De Los Santos, A. and Rickerson, R. J. and Siegmund, O. H. W.},
    month = apr,
    year = {2024},
    pages = {30},
}

@article{shen_resonant_1977,
    title = {Resonant frequency of a circular disc, printed-circuit antenna},
    volume = {25},
    issn = {1558-2221},
    url = {https://ieeexplore.ieee.org/document/1141643},
    doi = {10.1109/TAP.1977.1141643},
    number = {4},
    urldate = {2025-10-07},
    journal = {IEEE Transactions on Antennas and Propagation},
    author = {Shen, L. and Long, S. and Allerding, M. and Walton, M.},
    month = jul,
    year = {1977},
    pages = {595--596},
}

@article{marti_calibration_2001,
    title = {Calibration facility for solar wind plasma instrumentation},
    volume = {72},
    issn = {0034-6748},
    url = {https://doi.org/10.1063/1.1340020},
    doi = {10.1063/1.1340020},
    number = {2},
    urldate = {2025-10-02},
    journal = {Review of Scientific Instruments},
    author = {Marti, Adrian and Schletti, Reto and Wurz, Peter and Bochsler, Peter},
    month = feb,
    year = {2001},
    pages = {1354--1360},
}

@misc{noauthor_ieee_2011,
    title = {{IEEE} {Standard} for {Transitions}, {Pulses}, and {Related} {Waveforms}},
    url = {https://ieeexplore.ieee.org/document/6016198},
    doi = {10.1109/IEEESTD.2011.6016198},
    urldate = {2025-10-03},
    month = sep,
    year = {2011},
}

@inproceedings{fausch_advancing2_2025,
	title = {Advancing Lunar Exploration: The Neutral Gas Mass Spectrometer for Regolith and Exosphere Analysis},
	url = {https://ieeexplore.ieee.org/document/11068730},
	doi = {10.1109/AERO63441.2025.11068730},
	shorttitle = {Advancing Lunar Exploration},
	eventtitle = {2025 {IEEE} Aerospace Conference},
	pages = {1--14},
	booktitle = {2025 {IEEE} Aerospace Conference},
	author = {Fausch, Rico and Elsener, Hans Rudolf and Hofer, Lukas and Jost, Jürg and Lasi, Davide and Piazza, Daniele and Wurz, Peter},
	urldate = {2025-10-02},
	year = {2025},
}

@inproceedings{Fausch2026,
    title        = {A Novel Laser-ablation Resonance-ionization Mass Spectrometer for in Situ Dating of Lunar Rocks},
    author       = {Fausch, Rico and Aebi, A. and Alexander, A. and Anderson, F.S. and Fagan, A. and Ferguson, S. and Hanson, M. and Head, J. and Joy, K. and Klein, V. and Levine, J. and Osterman, S. and Pernet-Fisher, J. and Singh, V. and Tartese, R. and Teichmann, T. and Wurz, P. and Yant, M.},
    year         = {2026},
    eventtitle   = {2026 {IEEE} Aerospace Conference},
	booktitle    = {2026 {IEEE} Aerospace Conference},
    note         = {in press}
}

@inproceedings{Wurz2026,
    title        = {Prototype of a Laser-based Mass Spectrometer for In Situ Dating of Rocks on Planetary Surfaces},
    author       = {Wurz, P. and Wiesendanger, R. and Anderson, S.F. and Whitaker, T. and Tulej, M. and Fausch, R.},
    year         = {2026},
    eventtitle   = {2026 {IEEE} Aerospace Conference},
	booktitle    = {2026 {IEEE} Aerospace Conference},
    note         = {in press}
}

@article{wurz_fast_1996,
    title = {Fast microchannel plate detector for particles},
    volume = {67},
    issn = {0034-6748},
    url = {https://doi.org/10.1063/1.1146975},
    doi = {10.1063/1.1146975},
    number = {5},
    urldate = {2024-10-03},
    journal = {Review of Scientific Instruments},
    author = {Wurz, Peter and Gubler, Lukas},
    month = may,
    year = {1996},
    pages = {1790--1793},
}

@article{vorburger_mass_2024,
    title = {Mass {Spectrometer} {Experiment} for a {Uranus} {Probe}},
    volume = {220},
    issn = {1572-9672},
    url = {https://doi.org/10.1007/s11214-024-01096-9},
    doi = {10.1007/s11214-024-01096-9},
    language = {en},
    number = {6},
    urldate = {2024-10-02},
    journal = {Space Science Reviews},
    author = {Vorburger, Audrey and Wurz, Peter and Helled, Ravit and Mousis, Olivier},
    month = aug,
    year = {2024},
    pages = {64},
}

@article{schletti_fast_2001,
    title = {Fast microchannel plate detector with an impedance matched anode in suspended substrate technology},
    volume = {72},
    issn = {0034-6748},
    url = {https://doi.org/10.1063/1.1344601},
    doi = {10.1063/1.1344601},
    number = {3},
    urldate = {2024-10-03},
    journal = {Review of Scientific Instruments},
    author = {Schletti, Reto and Wurz, Peter and Scherer, Stefan and Siegmund, Oswald H.},
    month = mar,
    year = {2001},
    pages = {1634--1639},
}

@techreport{genolini_design_2001,
    title = {Design of the {Photomultiplier} {Bases} for the {Surface} {Detectors} of the {Pierre} {Auger} {Observatory}},
    number = {IPNO DR-01-010},
    institution = {Institut de Physique Nucléaire d'Orsay},
    author = {Genolini, B and Trung, T Nguyen and Pouthas, J and Lhenry-Yvon, I and Parizot, E and Suomijärvi, T},
    month = mar,
    year = {2001},
}

@article{wurz_new_1998,
    title = {New test facility for solar wind instrumentation},
    volume = {71},
    issn = {0018-0238},
    journal = {Helvetica Physica Acta},
    author = {Wurz, P. and Marti, A. and Bochsler, P.},
    year = {1998},
    pages = {23--24},
}

@article{wurz_impedancematching_1994,
    title = {Impedance‐matching anode for fast timing signals},
    volume = {65},
    issn = {0034-6748},
    url = {https://doi.org/10.1063/1.1144914},
    doi = {10.1063/1.1144914},
    number = {4},
    urldate = {2024-10-03},
    journal = {Review of Scientific Instruments},
    author = {Wurz, Peter and Gubler, Lukas},
    month = apr,
    year = {1994},
    pages = {871--876},
}

@article{wiley_timeflight_1955,
    title = {Time‐of‐{Flight} {Mass} {Spectrometer} with {Improved} {Resolution}},
    volume = {26},
    issn = {0034-6748},
    url = {https://doi.org/10.1063/1.1715212},
    doi = {10.1063/1.1715212},
    number = {12},
    urldate = {2024-08-08},
    journal = {Review of Scientific Instruments},
    author = {Wiley, W. C. and McLaren, I. H.},
    month = dec,
    year = {1955},
    pages = {1150--1157},
}

@article{yang_review_2025,
    title = {A review of the development of miniature mass spectrometry for planetary exploration},
    volume = {13},
    issn = {2772-5774},
    url = {https://www.sciencedirect.com/science/article/pii/S2772577425000540},
    doi = {10.1016/j.greeac.2025.100258},
    urldate = {2025-11-19},
    journal = {Green Analytical Chemistry},
    author = {Yang, Dong and Wu, Xiangkun and Hou, Keyong},
    month = jun,
    year = {2025},
    pages = {100258},
}

@article{vazquez_situ_2021,
    title = {In {Situ} {Mass} {Spectrometers} for {Applications} in {Space}},
    volume = {40},
    copyright = {© 2020 John Wiley \& Sons Ltd.},
    issn = {1098-2787},
    url = {https://onlinelibrary.wiley.com/doi/abs/10.1002/mas.21648},
    doi = {10.1002/mas.21648},
    language = {en},
    number = {5},
    urldate = {2025-11-19},
    journal = {Mass Spectrometry Reviews},
    author = {Vazquez, Timothy and Vuppala, Sinduri and Ayodeji, Ifeoluwa and Song, Linxia and Grimes, Nathan and Evans-Nguyen, Theresa},
    year = {2021},
    pages = {670--691},
}

@article{ren_review_2018,
    title = {A review of the development and application of space miniature mass spectrometers},
    volume = {155},
    issn = {0042-207X},
    url = {https://www.sciencedirect.com/science/article/pii/S0042207X18302458},
    doi = {10.1016/j.vacuum.2018.05.048},
    urldate = {2025-11-19},
    journal = {Vacuum},
    author = {Ren, Zhengyi and Guo, Meiru and Cheng, Yongjun and Wang, Yongjun and Sun, Wenjun and Zhang, Huzhong and Dong, Meng and Li, Gang},
    month = sep,
    year = {2018},
    pages = {108--117},
}

@inproceedings{schertenleib_ion-optical_2024,
    title = {Ion-{Optical} {Design} of a {Mass} {Spectrometer} for {Analyzing} {Complex} {Molecules} during {Fast} {Flybys}},
    url = {https://ieeexplore.ieee.org/abstract/document/10521245},
    doi = {10.1109/AERO58975.2024.10521245},
    urldate = {2025-11-20},
    booktitle = {2024 {IEEE} {Aerospace} {Conference}},
    author = {Schertenleib, Janis and Fausch, Rico G. and Wurz, Peter},
    month = mar,
    year = {2024},
    pages = {1--7},
}

@article{hohl_mass_1999,
    title = {Mass selective blanking in a compact multiple reflection time-of-flight mass spectrometer},
    volume = {188},
    issn = {1387-3806},
    url = {https://www.sciencedirect.com/science/article/pii/S1387380699000408},
    doi = {10.1016/S1387-3806(99)00040-8},
    number = {3},
    urldate = {2024-05-29},
    journal = {International Journal of Mass Spectrometry},
    author = {Hohl, M and Wurz, P and Scherer, S and Altwegg, K and Balsiger, H},
    month = jun,
    year = {1999},
    pages = {189--197},
}

@article{kurokawa_power_1965,
    title = {Power {Waves} and the {Scattering} {Matrix}},
    volume = {13},
    issn = {1557-9670},
    url = {https://ieeexplore.ieee.org/document/1125964},
    doi = {10.1109/TMTT.1965.1125964},
    number = {2},
    urldate = {2026-03-04},
    journal = {IEEE Transactions on Microwave Theory and Techniques},
    author = {Kurokawa, K.},
    month = mar,
    year = {1965},
    pages = {194--202},
}

@article{kuzminchuk_performance_2011,
    title = {Performance studies and improvements of a {Time}-of-{Flight} detector for isochronous mass measurements at the {FRS}-{ESR} facility},
    url = {http://dx.doi.org/10.22029/jlupub-9655},
    doi = {10.22029/jlupub-9655},
    language = {en},
    urldate = {2026-03-10},
    author = {Kuzminchuk, Natalia},
    year = {2011},
}

\end{document}